\documentclass{aa}  
\usepackage[switch]{lineno}
\usepackage{xcolor}
\usepackage{graphicx}
\usepackage{txfonts}
\usepackage[T1]{fontenc}
\usepackage{comment}
\usepackage{orcidlink}
\usepackage{hyperref}
\hypersetup{
  colorlinks,
  citecolor=blue,
  linkcolor=blue,
  urlcolor=blue}

\begin{document} 

   \title{Formation of black holes from He stars}

\titlerunning{Formation of Black Holes from He Stars}

   \author{Gang Long\inst{1,2,3}
	\and
	Bo Wang\inst{1,2,3}
	\and
	Philipp Podsiadlowski\inst{4,5}
	\and
	Dongdong Liu\inst{1,2,3}
	\and
	Yunlang Guo\inst{6}
	\and
	Shuai Zha\inst{1,2,3}
	\and
	Hanfeng Song\inst{7}
	\and
	Zhanwen Han\inst{1,2,3}
}
        
\institute{Yunnan Observatories, Chinese Academy of Sciences, Kunming 650216, China\\
	\email{wangbo@ynao.ac.cn}
	\and
	University of Chinese Academy of Sciences, Beijing 100049, China
	\and
	International Centre of Supernovae (ICESUN), Yunnan Key Laboratory, Kunming 650216, China
	\and
	London Centre for Stellar Astrophysics, Vauxhall, London, United Kingdom
	\and
	University of Oxford, St Edmund Hall, Oxford, OX14 AR, United Kingdom
	\and
	School of Astronomy and Space Science, Nanjing University, Nanjing 210023, China
	\and
	College of Physics, Guizhou University, Guiyang 550025, China}
\authorrunning{Gang Long et al}

% \abstract{}{}{}{}{} 
% 5 {} token are mandatory
  \abstract
 {Massive He stars are potential candidates of type Ib/c supernova (SN) progenitors. Understanding their final fates remains a key issue in astrophysics.  
 	In this work, we investigate the evolution of He stars with initial masses from $ 5\,M_{\odot}$ to $ 65\,M_{\odot}$, focusing on the presupernova (pre-SN) core structures to assess their explodability.  
 	Our simulations indicate that the final core structure is determined by the CO core mass and the central ${}^{12}\mathrm{C}$ mass fraction at the end of core He burning, affecting the properties of central C-burning and the locations of convective shells.
 	The location of the last convective C-burning shell sets the mass of the C-free core, constraining the iron core mass and compactness. 
 	We found that the final compactness and iron core mass exhibit non-monotonic behavior with initial mass, suggesting that the boundary between neutron star and black hole formation is not a simple mass threshold.
 	This is due to core C/Ne burning becoming neutrino dominated. This process drives stronger core contraction, ultimately increasing the iron core mass and the final compactness.
 	In contrast, earlier core Ne/O/Si ignition and shell mergers inhibit core contraction, reducing both the iron core mass and final compactness. We also discuss the effects of metallicity and overshooting on the pre-SN core structure. These factors potentially affect the explodability of progenitors.}


   \maketitle
%
  % context heading (optional)
  {}

%
%----------------------------------------------------------------
\section{Introduction}\label{sec:intro}
Most massive stars reside in close binaries and interact with companion stars during their lifetimes, and these interactions have a significant impact on both their evolution and core structures \citep{sana2012binary,langer2012presupernova,moe2017mind,laplace2021different,marchant2024evolution,Chen2024PrPNP}.
Binary interactions are expected to strip the hydrogen-rich envelopes to varying degrees, leaving behind a stripped-envelope star, such as a Wolf-Rayet star or a He star \citep{podsiadlowski1992presupernova,pols2002helium,langer2012presupernova,yoon2010type,song2016massive,laplace2020expansion,klencki2022partial}. These stars are potential candidates for progenitors of type Ib/Ic supernovae (SNe) \citep{crowther2007physical,mcclelland2016helium,yoon2017towards}. Thus, He stars are widely used as proxy models to study the evolution of stripped-envelope stars  \cite[e.g.,][]{brown2001formation,pols2002helium,yoon2012nature,yoon2017towards,woosley2019evolution,ertl2020explosion,dessart2020supernovae,aguilera2022stripped,aguilera2023stripped,Wang2026RAA}.
However, it remains uncertain which He stars successfully explode and which ones fail. Understanding their final fates remains a key issue in gravitational-wave astronomy.

The final fates of massive stars depend on their presupernova (pre-SN) core structure, which is associated with the properties of the subsequent neutrino-driven core-collapse SN (CCSN) explosion  \cite[e.g.,][]{o2011black,fryer2012compact,ugliano2012progenitor,Foglizzo2015Explosion,muller2016simple,sukhbold2016core,ertl2016two,couch2020simulating,burrows2021core,schneider2021pre,takahashi2023monotonicity,zha2023light,laplace2025s}. In this work, we focus on the pre-SN core structure to predict the explodability of progenitors, i.e., whether a star successfully explodes. A common method uses the compactness parameter $\xi_{2.5}$ to characterize the core structure and assess the explodability \citep{o2011black}. Higher values indicate a more compact core, suggesting that it is more likely to undergo a failed explosion and result in a (black hole) BH remnant \citep{o2011black,ugliano2012progenitor,sukhbold2014compactness,chieffi2020presupernova,schneider2021pre,temaj2024convective,laplace2025s}.
Another explodability criterion is based on the parameters $M_{4}$ and $\mu_4$, where $M_{4}$ typically corresponds to the Si/O shell interface and $\mu_4$ characterizes the density of the oxygen shell \citep{ertl2016two,ertl2020explosion,sukhbold2016core, muller2016simple, wang2022essential, boccioli2023explosion,heger2024black}.

All of these explodability criteria have indicated that the explodability of stars exhibits non-monotonic behavior with initial mass, characterized by islands of explodability \cite[e.g.,][]{o2011black,ertl2016two,ertl2020explosion,sukhbold2016core,sukhbold2018high,chieffi2020presupernova,patton2020towards,schneider2021pre,laplace2025s, Maltsev2025}. Several studies have investigated the physical mechanisms behind this behavior. For instance, the non-monotonic behavior of the final $\xi_{2.5}$ parameter with initial mass is influenced by the interaction between multiple C- and O-burning shells \citep{sukhbold2014compactness,chieffi2020presupernova}.  \cite{patton2020towards} and \cite{schneider2021pre} demonstrated that variations in the final compactness are correlated with the carbon-oxygen (CO) core mass and the residual ${}^{12} \mathrm{C}$ mass fraction at the end of core He burning. 
When neutrino energy loss exceeds the energy released by core C/Ne burning, stars experience a transition of core C/Ne burning from the convective to radiative regime, which is correlated with an increase in final compactness \citep{sukhbold2014compactness, sukhbold2020missing,chieffi2020presupernova, schneider2021pre, laplace2025s}. 
However, this non-monotonic behavior and the final fates of massive stars are highly sensitive to various uncertainties in their evolution, such as mass loss rates, metallicity, overshooting, and nuclear reaction rates, among others  \cite[e.g.,][]{smith2014mass,sukhbold2014compactness,renzo2017systematic,chieffi2021impact,aguilera2023stripped,temaj2024convective}.

In this work, we investigate the evolution of He stars and predict their final fates based on the pre-SN core structures. The paper is organized as follows. In Section \ref{sec:methods}, we describe the methods and physical assumptions in our numerical simulations. Section \ref{sec:results} presents the evolution and calculated results, followed by the discussion in Section \ref{sec:discussion} and the conclusions in Section \ref{sec:conclusion}.

\section{Methods and assumptions}\label{sec:methods}
We investigated the evolution of He stars using the \texttt{MESA} code \cite[r10398;][]{paxton2011modules,paxton2013modules,paxton2015modules,paxton2018modules,paxton2019modules} from the zero-age helium main sequence (ZaHeMS) to the onset of core collapse, which is defined as the point where the infall velocity at the iron core boundary reaches $1000 \, \mathrm{km} \, \mathrm{s}^{-1}$.
We constructed a grid of He star models following \cite{aguilera2022stripped}, with initial masses from $5\, M_{\odot}$ to $65\, M_{\odot}$, using mass increments of $\Delta M = 1\, M_{\odot}$ for $5$ to $30\, M_{\odot}$ and $\Delta M = 5\, M_{\odot}$ for $30$ to $65\, M_{\odot}$.
The models were computed with an initial metallicity of $Z=0.02$ and the \texttt{approx21\_cr60\_plus\_co56.net} nuclear reaction network, adopting the ${ }^{12} \mathrm{C}(\alpha, \gamma)^{16} \mathrm{O}$ reaction rate from \cite{angulo1999compilation}. The ${ }^{12} \mathrm{C}(\alpha, \gamma)^{16} \mathrm{O}$ reaction is one of the most important reactions in the evolution of massive stars, yet its rate is still highly uncertain \cite[e.g.,][]{sukhbold2014compactness,deboer2017c}. \cite{Long2025RAA} suggest that different rates for this reaction significantly affect the pre-SN core structure and explodability of He stars.

Convection was treated using the mixing length theory with a mixing length parameter $\alpha_{\text{MLT}} = 2.0$ \citep{bohm1958wasserstoffkonvektionszone}.
We applied the Ledoux criterion for convective boundaries and included semi-convective mixing with an efficiency parameter $\alpha_{\text{SEM}}=1.0$. During core He burning, we also included exponential overshooting with a parameter $f_{\rm ov}=0.01$ \cite[e.g.,][]{herwig2000,marchant2019pulsational,renzo2020predictions}. Additionally, we computed the models using the spatial and temporal resolution of \texttt{MESA} with \texttt{mesh\_delta\_coeff} = 0.6 and $\texttt{varcontrol\_target}$ = $10^{-5}$.

\subsection{Stellar wind mass loss} \label{sec:mass-loss}
To calculate the mass loss rate of He stars, we followed the prescription of \cite{yoon2017towards}, who systematically studied the mass loss rates for all subtypes of Wolf-Rayet (WR) stars (WNE, WC, and WO). These prescriptions are based on the work of \cite{tramper2016new} for WC and WO stars and \cite{hainich2014wolf} for WNE stars. The mass loss rates were calculated as follows. For WC/WO stars with a surface He mass fraction $Y_{\rm {S}} \leq 0.9$,
\begin{equation}\label{stellar_wind1}
	\dot{M}_{\mathrm{WC}}=f_{\mathrm{WR}}\left(\frac{L}{L_{\odot}}\right)^{0.85}\left(\frac{Z_{\text {init }}}{Z_{\odot}}\right)^{0.25} {Y_{\rm S}}^{0.44} 10^{-9.2} \frac{M_{\odot}}{\mathrm{yr}} .
\end{equation}
For WNE stars with $Y_{\text{S}}$ > 0.98,
\begin{equation}\label{stellar_wind2}
	\dot{M}_{\mathrm{WN}}=f_{\mathrm{WR}}\left(\frac{L}{L_{\odot}}\right)^{1.18}\left(\frac{Z_{\text {init }}}{Z_{\odot}}\right)^{0.6} \times 10^{-11.32} \frac{M_{\odot}}{\mathrm{yr}}.
\end{equation}
For 0.9 < $Y_{\text{S}}$ < 0.98, the mass loss rate we used is as follows,
\begin{equation}\label{stellar_wind3}
	\dot{M}=(1-f) \dot{M}_{\mathrm{WN}}+f \dot{M}_{\mathrm{WC}}
\end{equation}
with $f=\left(1-Z_{\text {init }}-Y\right) /\left(1-Z_{\text {init }}-0.9\right)$.\, We adopted the standard scaling factor $f_{\mathrm{WR}}=1.0$ and solar metallicity $Z_{\odot}$= 0.02.

\subsection{Explodability criterion} \label{sec:criterion}

The compactness parameter is often used as a criterion to characterize the core structure and assess the explodability of progenitors.
It is defined by \cite{o2011black} as

\begin{equation}\label{xi}
	\xi_M \equiv \frac{M / M_{\odot}}{R(M) / 1000 \mathrm{~km}}, \end{equation}
where $M$ is typically evaluated at a mass coordinate of $M = 2.5\,M_{\odot}$, and $ R(M)$ is the
corresponding radius. 
For progenitors with $\xi_{2.5}\lesssim0.45$, they are more likely to explode successfully, resulting in neutron stars (NSs), whereas at higher compactness values, they are more likely to collapse into BHs \citep{o2011black}. While \cite{ugliano2012progenitor} suggested that BHs may originate from progenitors with $\xi_{2.5} \gtrsim 0.35$, those with $\xi_{2.5} \lesssim 0.15$ are more likely to produce NSs.

However, the compactness parameter does not seem to be a reliable predictor of explodability, as both high- and low-compactness models can explode \cite[e.g.,][]{ertl2016two,muller2016simple, burrows2021core,wang2022essential,zha2023light,takahashi2023monotonicity,Burrows2020MNRAS,Burrows2023ApJ}. In order to estimate the fate after core collapse, we used the semi-analytic SN code of \cite{muller2016simple} to obtain the properties of successful neutrino-driven CCSN explosion, such as the explosion energy and the compact remnant mass. 
We adopted a maximum gravitational NS mass of 2 $M_{\odot}$, assuming a direct collapse into a BH for models with remnant masses exceeding this value \citep[see also][]{muller2016simple,schneider2021pre,schneider2024pre,temaj2024convective}. 
For comparison, we also used the two-parameter criterion to judge the explodability, which was proposed by \cite{ertl2016two}. Their criterion is based on $M_4$, the mass coordinate at which the specific entropy per nucleon equals four (typically corresponding to the Si/O interface), and $\mu_4$, the density gradient at that mass coordinate,
\begin{equation}\label{mu_4}
	\left.\mu_4 \equiv \frac{d m / M_{\odot}}{d r / 1000 \mathrm{~km}}\right|_{s=4}.
\end{equation}The $\mu_{4}$ and $M_{4}\mu_{4}$ are loosely correlated with the accretion rate and the accretion luminosity after the infall of the Si/O shell interface \citep{muller2016simple,ertl2016two,ertl2020explosion}. 
We used the relation $\mu_4 = 0.294 \mu_4 M_4 + 0.0468$ from the updated calibration model S19.8 to distinguish between explosion and non-explosion models (see \citealp{ertl2020explosion} for details).

\section{Results}\label{sec:results}

We investigated the evolution of the compactness parameter $\xi_{2.5}$ as a function of time until core collapse for a representative model. The pre-collapse structural properties of all models were subsequently characterized, from which their final fates were predicted. Finally, we explored the impact of metallicity and overshooting on the pre-SN core structure. Tables~\ref{parameter}--\ref{physical parameters} list the initial model configurations used in this work along with a summary of the properties at the onset of core collapse.
\subsection{A representative example}\label{sec:example}

\begin{figure*}
	\centering
	\includegraphics[width=1\textwidth]{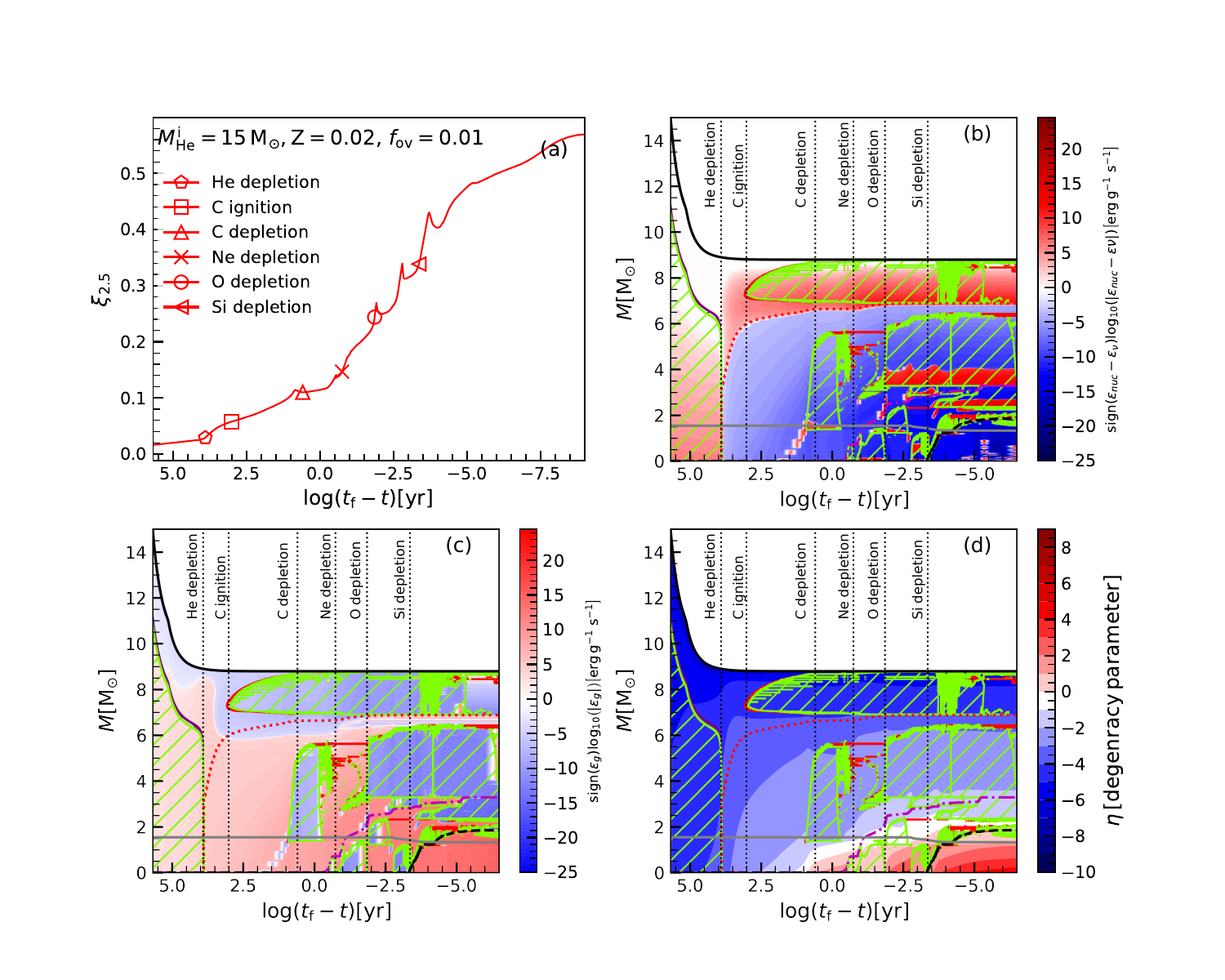}
	\caption{Evolutionary track of a 15 $M_{\odot}$ He star with solar metallicity ($Z = 0.02$) and overshooting parameter $f_{\mathrm{ov}} = 0.01$.
		Panel (a): Temporal evolution of the compactness parameter $\xi_{2.5}$ until seconds before core collapse. Key evolutionary phases are marked as follows: pentagram (central He depletion), square (central C ignition), triangle (central C depletion), cross (central Ne depletion), open circle (central O depletion), and left-pointing triangle (central Si depletion).
		Panel (b): Kippenhahn diagram from core He-burning to the onset of core collapse. The color bar shows the nuclear energy generation rate ($\epsilon_{\rm nuc}$, red) and neutrino loss rate ($\epsilon_{\nu}$, blue).
		The green, purple, and red-hatched regions represent the convective, overshooting, and semiconvective mixing regions, respectively. The gray solid indicate the classical Chandrasekhar mass, $M_{\rm Ch,0}$, as defined in Eq.$\,$\ref{eq:M_ch}. 
		The dashed black, dotted red, and dash-dotted magenta lines denote the iron, He-free ($M_{\text{He-free}}$), and C-free ($M_{\text{C-free}}$) cores, respectively.
        Vertical dotted lines mark specific depletion and ignition events.
		Panel (c): Same as panel (b), but the color represents the specific gravothermal energy, $\epsilon_{\text {grav }} \equiv-T ds / dt$ (red: contraction , blue: expansion).
		Panel (d): Same as panel (b), but the color represents the electron degeneracy parameter, $\eta \equiv \mu/k_{\rm B}T$, where $\mu$ is the chemical potential, $k_{\rm B}$ is the Boltzmann constant, and $T$ is temperature. $\eta \ll -1$: non-degenerate; $\eta \approx 0$: partial degeneracy; $\eta \gg 1$: strong degeneracy.}
	\label{fig:xi} 		
\end{figure*}

\begin{figure}
	\centering
	\includegraphics[width=1\columnwidth]{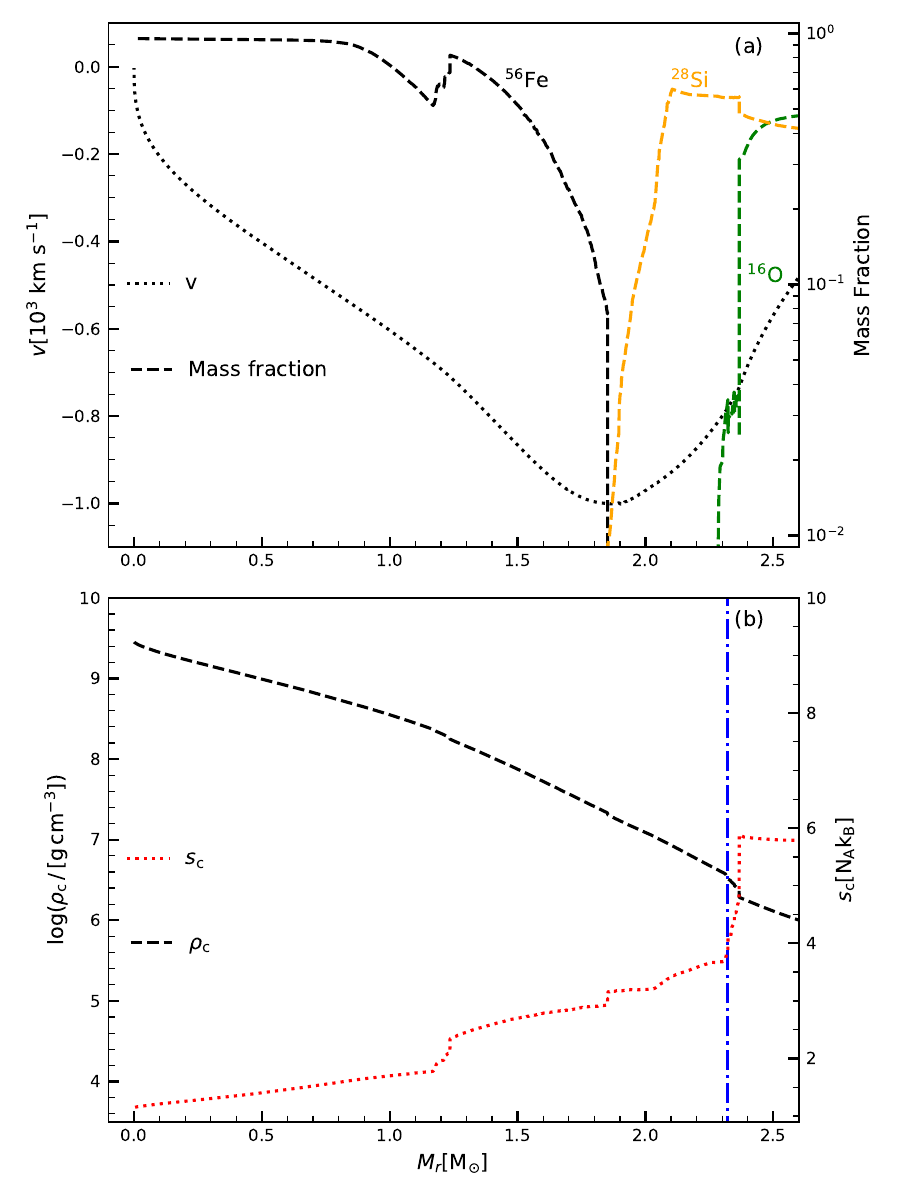}
	\caption{Same model as Fig.~\ref{fig:xi}. The figure shows the core structure of the progenitor at the onset of core collapse. Panel (a): Radial infall velocity (black dotted line, left axis) and mass fraction  profiles of $^{56}$Fe, $^{28}$Si, and $^{16}$O (dashed lines, right axis) versus the mass coordinate $M_{r}$. Panel (b): Radial profiles of the central density (dashed line) and specific entropy (dotted line) as functions of the mass coordinate $M_{r}$, where the iron core mass is 1.85 $M_{\odot}$. The dash-dotted blue line marks the mass coordinate where the central specific entropy $s_{\rm c}=4\, k_{\rm B}$, corresponding to the Si/O shell interface at $M_{r} = 2.32 \,M_{\odot}$.}
	\label{fig:oneset of collapse}
\end{figure}

 Fig.~\ref{fig:xi} shows the evolution of the $\xi_{2.5}$ parameter (panel a) and the stellar structure (panels b-d) of a $15\,M_{\odot}$ He star with $Z = 0.02$ and $f_{\rm ov} = 0.01$ from the ZaHeMS to the onset of core collapse. In this figure, key evolutionary phases are marked based on specific thresholds: Central C ignition corresponds to $T_{\rm c} \geq 8 \times 10^{8}$ K, while the depletion of fuel ($^{4}\rm He$, $^{12}\rm C$, $^{20}\rm Ne$, $^{16}\rm O$, and $^{28}\rm Si$) is defined as the point when the central mass fraction drops below $10^{-4}$.
The evolution of $\xi_{2.5}$ with stellar age is governed by the complex interplay between the various convective shells (i.e., C, Ne, and O shells) and the effects of electron degeneracy pressure \citep{sukhbold2014compactness, renzo2017systematic, chieffi2020presupernova}. As shown in Fig.~\ref{fig:xi}a, during core He burning, $\xi_{2.5}$ gradually increases because the convective He-burning core contracts overall. This contraction is displayed in Fig.~\ref{fig:xi}c by the specific gravothermal energy, defined as $\epsilon_{\text {grav }} \equiv-T ds / dt$, where $T$ is the temperature and $ds / dt$ is the time derivative of specific entropy; red regions indicate contraction, while blue regions indicate expansion.
At the end of core He burning, the values of $M_{\rm CO}$ and $X_{\rm C}$ are  7.51 $M_{\odot}$ and 0.254, respectively (see Table \ref{parameter}). 
Following core He depletion, $\xi_{2.5}$ continues to increase until central C depletion, as central C-burning proceeds radiatively (Fig.~\ref{fig:xi}b). This radiative process is driven by the reduction in $X_{\rm C}$, which lowers the nuclear energy generation rate below the neutrino energy loss rate.
After central C depletion, the significant decrease in the energy production rate induces contraction of both the core and the shells, which persists until the temperature is high enough to ignite shell C burning. The C-burning front also moves outward in mass from the center (see Fig. \ref{fig:xi}b). Notably, the burning front corresponds to the mass coordinate of the peak energy generation rate (indicated by the red regions in Fig. \ref{fig:xi}b), which also represents the base of the convective burning zone. The location of the burning front reflects the degree of core contraction that occurred beneath it \citep{laplace2025s}.
During the first convective C-burning shell (between core C depletion and Ne ignition), the $\xi_{2.5}$ value remains nearly constant (see Fig. \ref{fig:xi}a). This is because the C-burning front resides near, but below, $M_{\rm Ch, 0}$, with electron degeneracy pressure effectively resisting further core contraction (see Fig. \ref{fig:xi}d). To illustrate the degree of degeneracy, Fig.~\ref{fig:xi}d presents the electron degeneracy parameter, $\eta \equiv \mu/k_{\rm B}T$, where $\mu$ is the chemical potential, $k_{\rm B}$ is Boltzmann constant, and $T$ is temperature, distinguishing between regimes of non-degeneracy ($\eta \ll -1$), partial degeneracy ($\eta \approx 0$), and strong degeneracy ($\eta \gg 1$).

Following the onset of core Ne burning, the growth rate of $\xi_{2.5}$ increases substantially. Notably, the evolution of $\xi_{2.5}$ exhibits distinct oscillations after the end of core O burning (see the open circle in Fig.~\ref{fig:xi}a). These oscillations are driven by variations in the Si- and O-burning shells \citep{renzo2017systematic}. For example, as shown in Fig.~\ref{fig:xi}b, the first O-burning convective shell (ignited at log$(t_{\rm f} - t) \approx -2\, \rm yr$) is located within the mass coordinate used to evaluate the compactness. This shell releases a significant amount of energy, causing the outer layers to expand (see Fig.~\ref{fig:xi}c), which in turn reduces $\xi_{2.5}$. The compactness parameter $\xi_{2.5}$ ultimately reaches a value of 0.571. The final mass of the C-free core is $3.28 \, M_{\odot}$, which sets an upper limit on the growth of the Si-rich core and thus constrains the final iron core mass to $1.85\,M_{\odot}$ \citep[e.g.,][]{brown2001formation,schneider2021pre,laplace2025s}. As \cite{Fryer2002ApJ} suggested, the edge of the C-burning shell marks the boundary of the C-free core, and we defined the He-free and C-free core boundaries as the mass coordinates where $X_{\rm He} < 10^{-5}$ and $X_{\rm C} < 10^{-5}$, respectively \cite[see also][]{schneider2021pre}.

Fig.~\ref{fig:oneset of collapse} presents the core structure of the progenitor at the onset of core collapse.
Fig.~\ref{fig:oneset of collapse}a shows the infall velocity profile and the mass fraction profiles of $^{56}\rm Fe$, $^{28}\rm Si$, and $^{16}\rm O$ as a function of mass coordinate for the representative model at the onset of core collapse. These abundance curves depict the iron core boundary and the Si/O shell interface. As illustrated in Fig.~\ref{fig:oneset of collapse}b, the central density profile exhibits no significant discontinuity at the Si/O shell interface (see the dash-dotted blue line in Fig.~\ref{fig:oneset of collapse}b) at the onset of core collapse, suggesting a low likelihood of a successful explosion \citep{wang2022essential,boccioli2023explosion}. For $M_4 = 2.32$ and $\mu_4 = 0.16$ (see Fig.~\ref{fig:oneset of collapse}b), the criterion of \cite{ertl2016two} predicts no explosion. Similarly, the semianalytic approach of \cite{muller2016simple} also predicts that the star fails to explode, resulting in a gravitational remnant mass of $8.79\, M_{\odot}$.

\subsection{Pre-collapse structural characteristics of He stars}\label{sec:structure}

\begin{figure}
	\centering
	\includegraphics[width=1\columnwidth]{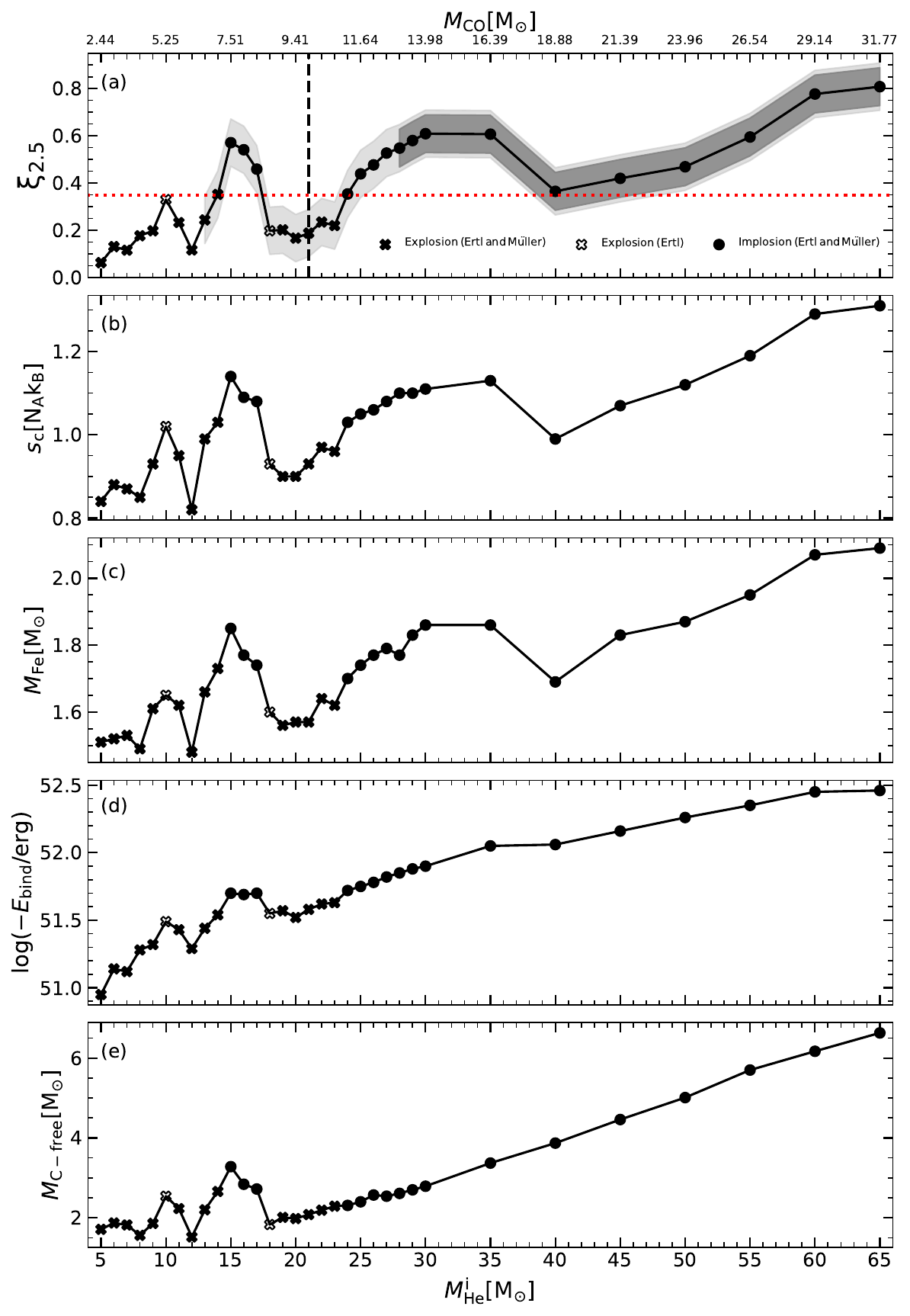}
	\caption{Key parameters at the onset of core collapse. Panel (a): Final compactness parameter $\xi_{2.5}$. Panel (b): Central specific entropy $s_c$. Panel (c): Iron core mass $M_{\rm Fe}$. Panel (d): Binding energy above the iron core $E_{\text{bind}}$. Panel (e): C-free core mass $M_{\rm C\text{-free}}$, all plotted as functions of the initial helium star mass $M_{\mathrm{He}}^{\mathrm{i}}$.
		The top axis shows the CO core masses of He stars at the end of core He burning. The light gray shading indicates radiative core C burning, while the darker gray shading represents radiative core Ne burning in the panel (a). The red dotted line at $\xi_{2.5} = 0.35$ separates models likely to explode (below the line) from those likely to implode (above the line). Black crosses indicate models predicted to explode according to both the explodability criteria of \cite{ertl2016two} and \cite{muller2016simple}. White crosses mark models predicted to explode by the \cite{ertl2016two} criterion but not the  \cite{muller2016simple} criterion. Black circles indicate models predicted to fail to explode and collapse into BHs according to both criteria. The black dashed vertical line marks the $M_{\mathrm{He}}^{\mathrm{i}} = 21\,M_{\odot}$ model ($M_{\rm CO} = 9.86\,M_{\odot}$), serving as a proxy for the $\sim 10\,M_{\odot}$ threshold for the upper limit of successful explosions suggested by \cite{patton2020towards}.}
	\label{fig:compactness} 		
\end{figure}

\begin{figure*}
	\centering
	\includegraphics[width=1\textwidth]{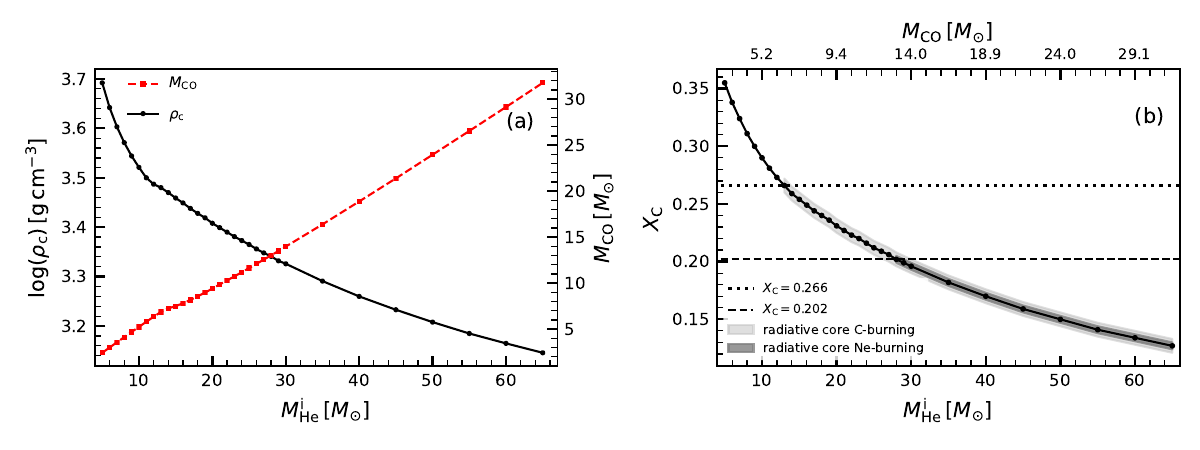}
	\caption{Core properties at the end of core He burning as a function of initial mass. Panel (a): Central density $\rho_{\mathrm{c}}$ and CO core mass $M_{\mathrm{CO}}$ at the end of core He burning, shown as a function of the initial He star mass $M_{\mathrm{He}}^{\mathrm{i}}$. Panel (b): Corresponding central ${}^{12}\rm C$ mass fraction $X_{\rm C}$ at the end of core He burning, with the top axis indicating the CO core mass $M_{\mathrm{CO}}$ for each $M_{\mathrm{He}}^{\mathrm{i}}$. Light and dark gray shaded regions denote models undergoing radiative core C and Ne burning, respectively. 
		The dotted and dashed lines represent the threshold ${}^{12}\rm C$ mass fractions separating convective and radiative burning regimes.}
	\label{fig:mass_fraction}
\end{figure*}

\begin{figure*}
	\centering
    \includegraphics[width=1\textwidth]{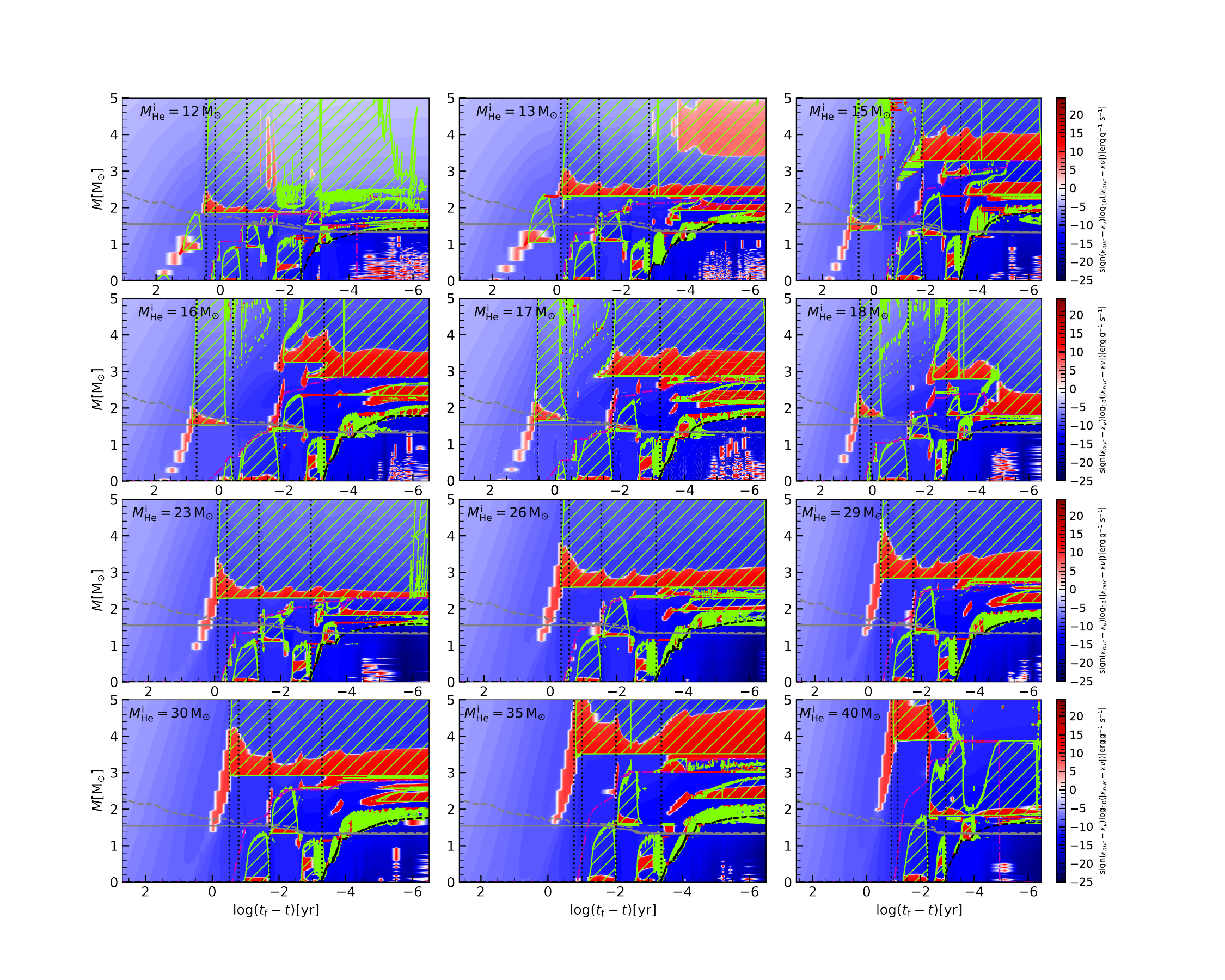}
	\caption{Same as Fig.~\ref{fig:xi}b, but for models with $M_{\mathrm{He}}^{\mathrm{i}}=12, 13, 15, 16, 17, 18, 23, 26, 29, 30, 35, \text{and}\, 40 \,M_{\odot}$, covering the evolution from core C-burning to the onset of core collapse. The gray dashed line indicates, $M_{\rm Ch}$, as defined by Eq. \ref{eq:M_ch}. The dotted vertical lines indicate, from left to right, the times of depletion of $^{12}$C, $^{20}$Ne, $^{16}$O, and $^{28}$Si in the core.}
	\label{fig:11-18kippen}
\end{figure*}

To characterize the pre-SN core structure and assess the explodability of stars, we focus on five key quantities at the onset of core collapse, including the compactness parameter $\xi_{2.5}$, central specific entropy $s_{\text{c}}$, iron core mass $M_{\text{Fe}}$, C-free core mass $M_{\rm C-free}$, and the binding energy above the iron core $E_{\text{bind}}$. 
Fig.~\ref{fig:compactness} shows these quantities as functions of initial mass, revealing consistent non-monotonic trends. These trends reflect the core's structural properties, as demonstrated in previous studies \citep[e.g.,][]{o2011black,ertl2016two, schneider2021pre, schneider2024pre, takahashi2023monotonicity, temaj2024convective, laplace2025s}.
Specifically, the final central entropy and the final iron core mass both follow trends similar to the final compactness parameter as a function of the initial mass. The relationship between the final central entropy (Fig.~\ref{fig:compactness}b) and the final iron core mass (Fig.~\ref{fig:compactness}c) can be interpreted in terms of the effective Chandrasekhar mass $M_{\text{Ch}}$ \citep{Timmes1996ApJ,woosley2002evolution,sukhbold2014compactness,schneider2024pre,laplace2025s}. 
The effective Chandrasekhar mass, $M_{\text{Ch}}$, is given by \cite{suwa2018minimum}:
\begin{equation}
	M_{\mathrm{Ch}}\approx M_{\rm Ch,0}\left[1+\left(\frac{s_{\mathrm{e}, \mathrm{c}}}{\pi Y_{\mathrm{e}, \mathrm{c}}}\right)^2\right],\label{eq:M_ch}
\end{equation}
where $M_{\rm Ch,0}=1.09 \left(\frac{Y_{\mathrm{e}, \mathrm{c}}}{0.42}\right)^2$, $Y_{\mathrm{e}, \mathrm{c}}$ is the central value of the electron fraction $Y_{\mathrm{e}}$, and $s_{\mathrm{e}, \mathrm{c}}$ is the central value of the electronic entropy per baryon $s_{\mathrm{e}}$, which is approximately one-third of the average central specific entropy $s_{\mathrm{c}}$ \citep{baron1990effect, Timmes1996ApJ}.
When the iron core reaches the Chandrasekhar mass $M_{\text{Ch}}$, core collapse is triggered. However, $M_{\text{Ch}}$ is sensitive to variations in $s_{\mathrm{c}}$ and $Y_{\mathrm{e}, \mathrm{c}}$ (see Eq. \ref{eq:M_ch}), with the latter remaining nearly constant across different progenitors \citep{sukhbold2014compactness}. Thus, $M_{\mathrm{Fe}}$ is closely correlated with $s_{\mathrm{c}}$ at the time of core collapse.

Furthermore, the final iron core is in an adiabatic state, with its entropy (Fig.~\ref{fig:compactness}b) following the polynomial relation $s_{\mathrm{c}} \propto \ln K \propto 2 \ln M + \ln R$ \citep{schneider2021pre}. 
This correlation further elucidates the connection between $s_{\mathrm{c}}$ and the mass-radius relation of the core, as reflected in the relationship between $s_{\mathrm{c}}$ and $\xi_{2.5}$ (Fig.~\ref{fig:compactness}a).
Additionally, we find that within the initial mass range of $5-30\,M_{\odot}$, both $E_{\text{bind}}$ and $M_{\rm C-free}$ follow non-monotonic trends, similar to those of $\xi_{2.5}$, $s_{\text{c}}$, and $M_{\text{Fe}}$. The iron core mass is constrained by the C-free core within this range \citep{brown2001formation, schneider2021pre, laplace2025s}. Beyond this range, both $E_{\text{bind}}$ and $M_{\rm C-free}$ increase approximately linearly.

Fig.~\ref{fig:mass_fraction} displays the central density, $\rho_{\mathrm{c}}$, CO core mass, $M_{\mathrm{CO}}$, and central ${}^{12}\rm C$ mass fraction, $X_{\mathrm{C}}$, at the end of core He burning for all models. Obviously, higher-mass He stars exhibit lower central densities, despite having a higher CO core mass (Fig.~\ref{fig:mass_fraction}a).  According to \cite{deboer2017c}, the molar abundance $Y(^{12}\mathrm{C})$ depends on the molar abundance of $^4\mathrm{He}$, the central density $\rho_{\rm c}$, and the corresponding reaction rates $\lambda_{3 \alpha}$ and $\lambda_{{}^{12} \mathrm{C}(\alpha, \gamma)^{16} \mathrm{O}}$, as shown in the rate equation
\begin{equation}
	\frac{d Y\left({ }^{12} \mathrm{C}\right)}{d t}=\frac{1}{3!} Y^3\left({ }^4 \mathrm{He}\right) \cdot {\rho_{\rm c}}^2 \cdot \lambda_{3 \alpha}-Y\left({ }^{12} \mathrm{C}\right) \cdot Y\left({ }^4 \mathrm{He}\right) \cdot \rho_{\rm c} \cdot \lambda_{{}^{12} \mathrm{C}(\alpha, \gamma)^{16} \mathrm{O}}.
	\label{eq:carbon_mass_fraction}
\end{equation}
This equation indicates that carbon production is favored at high densities. Therefore, higher-mass He stars exhibit lower central ${}^{12}\mathrm{C}$ mass fractions (see Fig.~\ref{fig:mass_fraction}b). At the end of core He burning, $X_{\mathrm{C}}$ and $M_{\mathrm{CO}}$ set the initial conditions for the subsequent burning stages, which have important consequences for the final core structure \cite[e.g.,][]{sukhbold2014compactness,chieffi2020presupernova,sukhbold2020missing, patton2020towards, schneider2021pre,laplace2025s}. Specifically, when the energy generation rate $\epsilon_{\mathrm{nuc}} \sim X_{\mathrm{C}}^2 \rho T^{23}$ (where $\rho$ and $T$ are density and temperature near the center) exceeds the neutrino cooling rate $\epsilon_{\nu} \sim T^{12} \rho^{-1}$, the burning mode becomes convective \citep{sukhbold2020missing}.
As shown in Fig.~\ref{fig:mass_fraction}b, when the value of $X_{\mathrm{C}}$ drops below approximately 0.266, core C burning phase transitions from convective to radiative. Similarly, when $X_{\mathrm{C}}$ falls below 0.202, core Ne burning phase also shifts to radiative. These transitions correspond to the gray and dark shaded regions in Fig.~\ref{fig:compactness}a. To understand these non-monotonic trends, we investigated the key mechanisms that shape the final compactness and core structure, which are analyzed in detail below. As shown in Fig.~\ref{fig:11-18kippen}, the Kippenhahn diagrams illustrate the evolution from core C burning to the onset of core collapse.
\subsubsection{Increase in final compactness: Core C-burning transition stage}
For the models with $M_{\mathrm{He}}^{\mathrm{i}} = 12-15\, M_{\odot}$, the final compactness $ \xi_{2.5} $ increases with mass (Fig.~\ref{fig:compactness}a), peaking at $ \xi_{2.5} = 0.571 $ for $M_{\mathrm{He}}^{\mathrm{i}} = 15\,M_{\odot}$. This trend is driven by the decrease in central density and carbon abundances with increasing initial mass (Fig.~\ref{fig:mass_fraction}), which triggers the transition from convective to radiative core C burning.
Fig.~\ref{fig:11-18kippen} shows the Kippenhahn diagrams that compare the evolution of the $12\, M_{\odot}$, $13 \, M_{\odot}$, and $15 \, M_{\odot}$ models.

The $12 \,M_{\odot} $ model exhibits a tiny convective C-burning core, while the $13 \, M_{\odot}$ and $15 \, M_{\odot}$ models exhibit radiative C-burning cores. This transition indicates that core C burning becomes neutrino-dominated, triggers stronger core contraction, and drives the outward progression of the C-burning front. After the convective or radiative C-burning core exhausts its fuel, the subsequent core contraction triggers the formation of the first convective C-burning shell. The C-burning front reaches the base of this shell at mass coordinates of 0.82 $M_{\odot}$, 1.03 $M_{\odot}$, and 1.37$\, M_{\odot}$ for the $12\, M_{\odot}$, $13 \, M_{\odot}$, and $15\,M_{\odot}$ models, respectively. As these C-burning fronts move further out in mass, it indicates stronger core contraction \citep{laplace2025s}.
For example, the $M_{\mathrm{He}}^{\mathrm{i}}=15 \, M_{\odot}$ model shows a more extended region of electron degeneracy compared to the other two models during the first convective C-burning shell (see Fig.~\ref{fig:He_12-18_eta_kippen}). Since the C-burning front in these models remains below $ M_{\rm Ch, 0}$ (see the gray solid lines in Fig.~\ref{fig:11-18kippen}), electron degeneracy pressure counteracts core contraction, thereby delaying the ignition of Ne- and O- burning \citep{sukhbold2014compactness,laplace2025s}. In the $15 \, M_{\odot}$ model, core Ne- and O-burning ignite sequentially after the first C-burning shell depletes its fuel, further delaying the outward progression of the C-burning front until O-burning ends. Subsequent core contraction triggers a second convective C-burning shell. The C-burning front moves out to a mass coordinate of $3.28\, M_{\odot}$ and remains there until the end of evolution, ultimately producing a large C-free core of $3.28\, M_{\odot}$ (indicated by the dash-dotted magenta line in Fig.~\ref{fig:11-18kippen}).
This results in a $1.88 \, M_{\odot}$ iron core and a compactness value of 0.57.

 In the $12\,M_{\odot}$ and $13\,M_{\odot}$ models, the second convective C-burning shell is triggered immediately after core C depletion (see the first dashed line in the top row of Fig.~\ref{fig:11-18kippen}) and remains at the same mass coordinate until collapse, which yields smaller C-free cores of $ 1.86 \, M_{\odot} $ and $ 2.20 \, M_{\odot} $, respectively. This suppresses Si-rich core growth, leading to smaller iron cores and reduced compactness.

\subsubsection{Decrease in final compactness: Convective shell merger stage}
For the models with $ M_{\mathrm{He}}^{\mathrm{i}} = 15-18\, M_{\odot}$, the final compactness gradually decreases (see Fig.~\ref{fig:compactness}a). This trend is primarily driven by earlier core Ne and O ignition as well as shell mergers. As shown in Fig.~\ref{fig:11-18kippen}, central C-burning in these models proceeds radiatively and is dominated by neutrino losses. After core C depletion (see the first dashed line in Fig.~\ref{fig:11-18kippen}), the first convective C-burning shell forms, and the C-burning front gradually shifts outward in mass coordinate.
For example, the front extends to approximately $1.37\, M_{\odot}$ in the $15\,M_{\odot}$ model, compared to $1.81\,M_{\odot}$ in the $18\, M_{\odot}$ model. 
The C-burning front approaches or exceeds $M_{\rm Ch,0}$ during the first convective C-burning shell. This triggers a rapid contraction of the partially degenerate core, accelerating the earlier ignition of Ne burning before the first covective C-burning shell is fully depleted \citep{laplace2025s}. The earlier ignition of core Ne and O enhances core luminosity, suppress core contraction, and slows the progression of the C-burning front. After core O depletion (see the third dashed line in Fig.~\ref{fig:11-18kippen}), further core contraction ignites the second C-burning convective shell, shifting the location of the C-burning front to $3.28\, M_{\odot}$, $3.27\, M_{\odot}$, $2.87\, M_{\odot}$, and $2.80\, M_{\odot}$ for the $15\, M_{\odot}$, $16\, M_{\odot}$, $17\, M_{\odot}$, and $18\, M_{\odot}$ models, respectively.

After core Si burning (see the last dashed line in Fig.~\ref{fig:11-18kippen}), the mass coordinate of the C-burning front in the $15\,M_{\odot}$ model remains constant at $3.28 \,M_{\odot}$. However, for models with $M_{\mathrm{He}}^{\mathrm{i}} = 16-18\,M_{\odot}$, its mass coordinate declines to varying degrees. This behavior is attributed to a substantial increase in both the nuclear energy generation and the entropy within the Ne/O-burning shells. When the entropy of these shells exceeds that of the C-burning shell above, a shell merger is triggered. Following the merger, the burning front moves to a lower mass coordinate and forms a larger convective shell, causing the overlying layers to expand \citep{sukhbold2014compactness,sukhbold2018high,collins2018properties,Davis2019MNRAS,laplace2025s}.
In the $16\,M_{\odot}$ and $ 17\,M_{\odot}$ models, a merger of the Ne shell with the C shell above occurs after core Si ignition. The burning front then moves inward to mass coordinates of $2.84\,M_{\odot}$ (at log$(t_{\rm f} - t) \approx -3.4\,\rm yr$) and $2.72\,M_{\odot}$ (at log$(t_{\rm f} - t) \approx -3.57\,\rm yr$), respectively, which are consistent with the masses of their C-free cores.
In the $18\,M_{\odot}$ model, in addition to experiencing the merger of the Ne and C shells, the O shell also merges with the Ne and C shells above. This process causes the mass coordinate of the burning front to decrease to $1.83 \, M_{\odot}$, corresponding to the mass of a C-free core (see the magenta dashed line in Fig.~\ref{fig:11-18kippen}), which leads to a smaller iron core and a corresponding reduction in $\xi_{2.5}$.

\subsubsection{Reincrease in final compactness: Core Ne-burning transition stage}\label{sec:Ne-burning}

For the models with $M_{\mathrm{He}}^{\mathrm{i}} = 23-30 \, M_{\odot} $, the final compactness gradually increases again, reaching a second peak of 0.60 at $M_{\mathrm{He}}^{\mathrm{i}} = 30 \,M_{\odot}$, which corresponds to $M_{\rm CO} \approx 13.98 \,M_{\odot}$ (see also Fig.~\ref{fig:compactness}a). 
With increasing initial mass, both the residual C mass fraction after central He burning depletion and the Ne mass fraction generated via the ${ }^{12} \mathrm{C}\left({ }^{12} \mathrm{C}, \alpha\right){ }^{20} \mathrm{Ne}$ reaction decrease. This drives a transition in the nature of core Ne burning, as well as in the number and size of Ne-burning shells (e.g., from multiple small zones to a few tiny ones, eventually transitioning to a fully radiative core Ne burning, see in the $23\,M_{\odot}$, $26\,M_{\odot}$, and $29\,M_{\odot}$ models in Fig.~\ref{fig:11-18kippen}).

As neutrino losses increase and the core contracts further, radiative Ne burning cannot effectively suppress the progression of the C-burning front. Furthermore, the increase in entropy from the Ne-burning shell is insufficient to trigger a merger with the C-burning shell above. Thus, the C-burning front continues to move outward in mass coordinates until it ignites a convective C-burning shell, maintaining this state until the onset of core collapse. In the $23\,M_{\odot}$, $26\,M_{\odot}$, and $29\,M_{\odot}$ models, the C-burning front reaches $2.29\,M_{\odot}$, $2.57\,M_{\odot}$, and $2.83\,M_{\odot}$, respectively. 
Since the mass coordinate of the C-burning front in the $26\,M_{\odot}$ and $29\,M_{\odot}$ models exceeds the mass coordinate used for the compactness parameter, this leads to a gradual increase in their final compactness.
Additionally, the C-free core nearly reaches the C-burning front in the $23\,M_{\odot}$ and $26\,M_{\odot}$ models, while in the $29\,M_{\odot}$ model, it reaches $2.70\,M_{\odot}$, slightly below the corresponding C-burning front.
These differences in the location of the C-burning front and the C-free core ultimately lead to a systematic increase in both the iron core mass and final compactness (e.g., in the $23\,M_{\odot}$, $26\,M_{\odot}$, and $29\,M_{\odot}$ models, the iron core masses are $1.62\,M_{\odot}$, $1.77\,M_{\odot}$, and $1.83\,M_{\odot}$, with corresponding final compactness values of $0.220$, $0.477$, and $0.580$, respectively).

\subsubsection{Subsequent decrease in final compactness: Core O- and Si-burning dominance stage}

For the models with $M_{\mathrm{He}}^{\mathrm{i}} = 30-40 \, M_{\odot} $, the final compactness decreases gradually once again (see also Fig.~\ref{fig:compactness}a). 
In these models, both C and Ne burn radiatively in the center (see Fig.~\ref{fig:11-18kippen}).
Notably, the location of their last C-burning convective shell significantly exceeds the mass coordinate used for the compactness parameter, thereby reducing the constraints on the final iron core mass and compactness. Consequently, the core O and Si burning, along with their convective shells, become progressively more important for the subsequent evolution.

In the 30$\,M_{\odot}$ model, the mass coordinate at the base of the first convective O-burning shell is close to $M_{\rm Ch,0}$ (see Fig.~\ref{fig:11-18kippen}). However, in the 35$\,M_{\odot}$ and 40$\,M_{\odot}$ models, it exceeds both $M_{\rm Ch,0}$ and $M_{\rm Ch}$, which accelerates core contraction and triggers early Si burning. As shown in Fig.~\ref{fig:11-18kippen}, core Si ignition occurs before the depletion of the first convective O-burning shell in the 35$\,M_{\odot}$ and 40$\,M_{\odot}$ models. Additionally, in the 40$\,M_{\odot}$ model, a larger convective shell forms and persists until the end of evolution. This delays the progression of the O-burning front in mass coordinate (e.g., the O-burning front reaches $2.22\,M_{\odot}$, $2.30\,M_{\odot}$, and $1.82\,M_{\odot}$ in the $30\,M_{\odot}$, $35\,M_{\odot}$, and $40\,M_{\odot}$ models, respectively). Consequently, this results in the formation of iron cores with smaller masses and reduced compactness.

\subsection{Final fates and compact-remnant masses} \label{sec:remnants}

We used the semianalytic approach of  \cite{muller2016simple} and the criterion of \cite{ertl2016two} to estimate the final fate after core collapse. As shown in Fig.~\ref{fig:compactness}a, using these methods, He stars with initial masses $M_{\mathrm{He}}^{\mathrm{i}} \leq 14\,M_{\odot}$ ($M_{\rm CO} \leq 7.27\,M_{\odot}$) are predicted to explode, except for the model with $M_{\mathrm{He}}^{\mathrm{i}} = 10\,M_{\odot}$ ($M_{\rm CO} = 5.25\,M_{\odot}$), which is predicted to explode according to the criterion of \cite{ertl2016two}, but not according to \cite{muller2016simple}.
At higher masses, the $M_{\mathrm{He}}^{\mathrm{i}} = 18\,M_{\odot}$ star is predicted to explode according to \cite{ertl2016two} but collapse following \cite{muller2016simple}. Both methods consistently predict that stars with $M_{\mathrm{He}}^{\mathrm{i}} \geq 24\,M_{\odot}$ ($M_{\rm CO} \geq 11.19\,M_{\odot}$) will fail to explode and instead directly collapse to form BHs. Although the compactness parameter alone is insufficient to determine whether a star will undergo a successful neutrino-driven SN explosion \citep[e.g.,][]{wang2022essential,Burrows2023ApJ}, our analysis suggests that for models with $\xi_{2.5} > 0.35$, both the semianalytic approach of \cite{muller2016simple} and the criterion of \cite{ertl2016two} predict that the star will fail to explode and is more likely to collapse into a BH remnant. This is illustrated in Fig.~\ref{fig:compactness}a, where the red dashed line at $\xi_{2.5} = 0.35$ separates models that may explode (below the line) from those that are more likely to collapse into a BH (above the line).

\begin{figure*}
	\centering
	\includegraphics[width=1\textwidth]{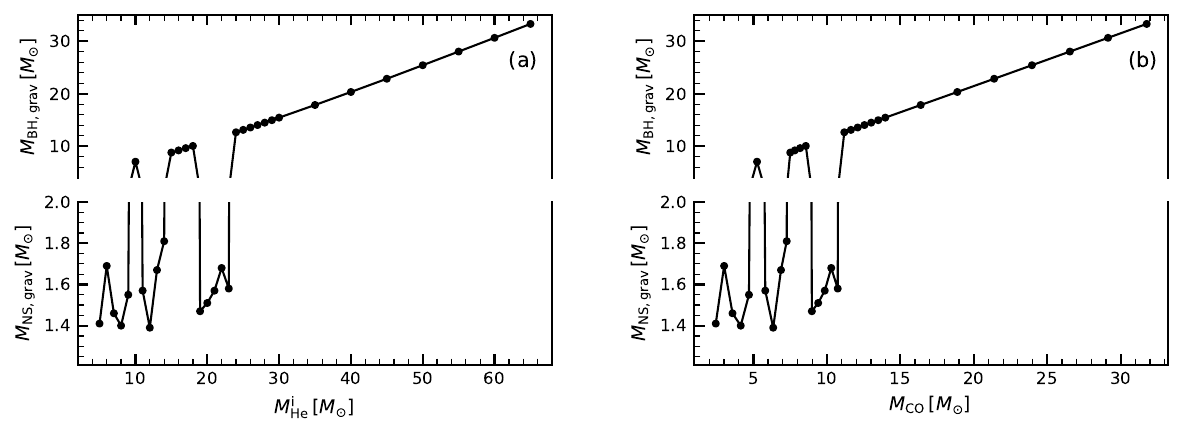}
	\caption{Gravitational masses of NSs, $M_{\rm{NS,grav}}$, and BHs, $M_{\rm{BH,grav}}$, as a function of both the $M_{\mathrm{He}}^{\mathrm{i}}$ in panel (a) and the $M_{\rm CO}$ in panel (b) for models with initial masses ranging from $5-65\, M_{\odot}$. The gravitational masses are calculated using the semianalytic approach of \cite{muller2016simple}.}
	\label{fig:Gravitational mass}
\end{figure*}

 Fig.~\ref{fig:Gravitational mass} shows the core-collapse outcomes for all models using semianalytic approach of \cite{muller2016simple}, presenting the mass distributions of compact remnants as functions of $M_{\mathrm{He}}^{\mathrm{i}}$ and $M_{\rm CO}$. The remnant mass distribution exhibits a non-monotonic trend with both $M_{\mathrm{He}}^{\mathrm{i}}$ and $M_{\rm CO}$, similar to the behavior of the compactness parameter. 
 This mapping is consistent with \cite{Patton2022MNRAS}, who predicted alternating regions of successful explosion and implosion based on the pre-SN core structure, rather than a continuous distribution. For instance, \cite{Patton2022MNRAS} identified a prominent peak in the BH mass function around $\sim 10\,M_{\odot}$ followed by a frequency drop above $\sim 11.5\,M_{\odot}$.
 Physically, this mapping corresponds to the correlation we observed in Fig.~\ref{fig:compactness} between BH formation and models with high final compactness, which are characterized by elevated binding energy, iron-core mass, carbon-free core, and central entropy. As noted by \citet{schneider2021pre}, all these key quantities exhibit non-monotonic behavior with $M_{\rm CO}$, which determines the pattern of NS and BH formation. Specifically, \citet{schneider2021pre} found that Case A and B stripped stars have higher $X_{\rm C}$ than single and Case C stripped stars at the same $M_{\rm CO}$. This difference alters the transition from convective to radiative core C/Ne burning, leading to a systematic shift of the island of explodability in $M_{\rm CO}$ (e.g., the first compactness peak shifts from $\sim 7 \,M_{\odot}$ for single and Case C stripped stars to $\sim 8  \,M_{\odot}$ for Case A and B stripped stars). Our models exhibit a peak at $M_{\rm CO} \approx 7.51  \,M_{\odot}$, which is higher than the value for single and Case C stripped stars. Thus, despite variations in the location of the islands of explodability across different systems, the underlying physical mechanism driving the non-monotonic behavior remains robust across different studies.

The predicted mass distributions of compact objects plays a critical role in subsequent binary evolution. When the primary star collapses into either a white dwarf (WD), NS, or BH as the first compact object, the subsequent evolution of the secondary star and its interaction with the first compact object can produce various types of compact binary systems. In particular, when He stars with $M_{\mathrm{He}}^{\mathrm{i}} \lesssim 3.2 \,M_{\odot}$ form WDs or NSs via electron-capture SNe \citep{woosley2019evolution}. The subsequent evolution of these systems, influenced by the properties of the companion and binary interactions, can result in double WDs, NS-WD binaries, or double NSs systems \citep[e.g.,][]{ tauris2013evolution,tauris2015ultra,tauris2017formation,wang2020formation,wang2022formation,guo2024electron}.
In our models, He stars with initial masses of 5--14 $M_{\odot}$ and 18--23 $M_{\odot}$ are expected to successfully explode and leave NS remnants as the first compact objects, while those with masses of 15 to 17 $M_{\odot}$ and $\geq$ 24 $M_{\odot}$ fail to explode and directly collapse to form BHs.
If the natal kick velocity is insufficient to disrupt the system, the resulting binary configuration comprises a compact object and an OBe-type star. 
These surviving systems ultimately evolve into observable high-mass X-ray binaries \citep[e.g.,][]{remillard2006x,kretschmar2019advances}, which are considered strong candidates for BH-BH or BH-NS progenitor systems \citep{belczynski2013cyg,neijssel2021wind}.

\begin{table*}
	\caption{Dependence of pre-SN properties on physical parameters.}\label{physical parameters}
	\begin{center}
		\begin{tabular}{
				l@{\hspace{2.5mm}}c@{\hspace{2.5mm}}c@{\hspace{2.5mm}}
				c@{\hspace{2.5mm}}c@{\hspace{2.5mm}}c@{\hspace{2.5mm}}
				c@{\hspace{2.5mm}}c@{\hspace{2.5mm}}c@{\hspace{2.5mm}}
				c@{\hspace{2.5mm}}c@{\hspace{2.5mm}}c@{\hspace{2.5mm}}
				c@{\hspace{2.5mm}}c@{\hspace{2.5mm}}c@{\hspace{2.5mm}}
				c@{\hspace{2.5mm}}c@{\hspace{2.5mm}}c@{\hspace{2.5mm}}
			}
			\hline\hline
			
			$M_{\text{He}}^{\text{i}}$  & $Z$&$f_{\rm {ov}}$& $M_{\rm CO}$&  $X_{\text{C}}$ &$M_{\text{C-free}}$ & $\xi_{2.5}$ &$\mu_{4}$&$M_{4}$& $s_{\text{c}}$ & $M_{\rm Fe}$ &$M_{\rm f}$&  log$(-E_{\text{bind}})$ & $E_{\text{exp}}$ & $\text{Fate}$  &$M_{\text{rm,grav}}$
			\\
			$\left[M_{\odot}\right]$ & &  & $\left[M_{\odot}\right]$ && $\left[M_{\odot}\right]$ &  &    &  &$\left[N_{\mathrm{A}} k_{\mathrm{B}}\right] $ &$\left[M_{\odot}\right]$  &$\left[M_{\odot}\right]$  &$\left[\text{erg}\right] $ & $\left[10^{51}\text{erg}\right] $  & &$\left[M_{\odot}\right]$ \\
			\hline
			15.0 & 0.0088   &   0.01 &  9.25&   0.247     &    1.92   &  0.145  &  0.05 &   1.67 &    0.87 &      1.56 & 11.5   &  51.49 &   0.24 & NS &1.46 \\		
			15.0 &  0.02   &   0.01 & 7.51 &0.254 & 3.28  &   0.571 &   0.16 & 2.32 &  1.14& 1.85 &8.79  &   51.70& $\cdots$ & BH & 8.79 \\												
			15.0 &  0.03   &   0.01 &    6.45 & 0.259    &   2.07    &       0.163 &   0.07 &    1.77 &  0.92 &    1.54 &      7.82 &   51.58 &0.56 & NS & 1.54  \\		
			15.0 &  0.04   &   0.01 &  5.81 & 0.262       &    2.16   &       0.189 &   0.08 &   1.82 &  0.95&      1.60  &  7.14 &   51.56 &0.77 & NS & 1.58 \\					\hline	    
			15.0 &  0.02   &   0.0 & 7.70 &  0.253     &     3.35  &        0.597  &   0.17 &    2.37 &  1.13&   1.89  &   8.96  &   51.72 & $\cdots$& BH & 8.96 \\	
			15.0 &  0.02   &   0.016 &  7.43 & 0.255     &   3.07    &        0.535 &   0.15 &    2.29&  1.12 & 1.85 &  8.70&   51.67 &$\cdots$ & BH & 8.70 \\		
			15.0 & 0.02   &   0.027 & 7.34 & 0.256      &   2.67   &   0.444 &   0.13 &    2.14 &  1.08 &   1.79 &   8.57&   51.61&$\cdots$& BH & 8.57 \\	    
			
			\hline
		\end{tabular}
	\end{center}
	{\bf Notes:} Same as Table \ref{parameter}, but for the representative model $M_{\text{He}}^{\text{i}}=15 \, M_{\odot}$, with variations in the physical parameters. We used different metallicities, $Z = 0.0088, 0.02, 0.03, 0.04$, and explored a range of convective overshooting parameters, $f_{\rm ov} = 0.0, 0.01, 0.016, 0.027$. The case with $Z = 0.02$ and $f_{\rm ov} = 0.01$ serves as the fiducial model.
\end{table*}

\section{Discussion} \label{sec:discussion}

\subsection{Comparison with previous studies} \label{sec:comparison}

In this work, we adopted He stars as the starting point to study the evolution of stripped-envelope stars. These models provide a physically motivated proxy for the progenitors of Type Ib/c SNe, which are expected to form through either strong stellar winds or binary interactions
 \cite[e.g.,][]{pols2002helium,crowther2007physical,yoon2017towards,woosley2019evolution,laplace2021different,marchant2024evolution}. Compared to evolving models from the zero-age main sequence (ZAMS), this approach reduces the uncertainties associated with mass loss during the red supergiant phase and binary interactions \cite[e.g.,][]{smith2014mass,renzo2017systematic,marchant2024evolution}. 
 Furthermore, in contrast to bare CO core models \cite[e.g.,][]{patton2020towards} which lack the He-shell burning phase and the steep pressure gradient at the boundary of the CO core, He star models allow for a self-consistent treatment of these features.
 As noted by \cite{patton2020towards}, bare CO cores tend to be denser in the center and cooler in the outer parts compared to equivalent cores within He stars. Such structural differences can shift the transition from convective to radiative C burning, thereby altering the islands of explodability \cite[e.g.,][]{sukhbold2014compactness}. However, using He stars as a proxy for stripped-envelope stars neglects the residual hydrogen envelope typically retained after binary interactions \cite[e.g.,][]{laplace2021different,aguilera2022stripped}. Even a trace amount of hydrogen can alter post-interaction evolution, most notably by triggering significant radial expansion \cite[e.g.,][]{Gilkis2019MNRAS,laplace2020expansion,dessart2020supernovae}.

Several works studied the evolution of He stars as a proxy for stripped-envelope stars, primarily focusing on the nature of their explosion \cite[e.g.,][]{woosley2019evolution,ertl2020explosion,dessart2020supernovae,aguilera2022stripped,aguilera2023stripped}.
In this work, we explored the pre-SN core structures of He stars by focusing on several quantities, including final compactness, iron core mass, central entropy, C-free core mass, and binding energy and their correlation with explodability.
Although no clear compactness threshold distinguishes between successful and unsuccessful explosions \cite[e.g.,][]{Burrows2020MNRAS,takahashi2023monotonicity}, we found that He stars with a final compactness of $\xi_{2.5} > 0.35$ are more likely to form BHs (see Fig. \ref{fig:compactness}). This result is consistent with that of \cite{aguilera2023stripped}. Moreover, \cite{aguilera2023stripped} suggested that most He stars with initial masses $\sim 35-40 \rm \,M_{\odot}$ will produce successful explosions. However, we found that most He stars with initial masses below 24 $M_{\odot}$ were expected to undergo successful explosions. 
These differences in the predicted outcomes arise from the different values adopted for the mass-loss rate factor $f_{\rm WR}$; we used $f_{\rm WR} = 1$, while their models used $f_{\rm WR} = 1.58$.

In addition, \cite{ertl2020explosion} investigated the collapse of solar-metallicity He star models of \cite{woosley2019evolution}, using a neutrino-driven hydrodynamics SN code. These models adopted different mass-loss rates, $f_{\rm WR} = 1.0$ and $f_{\rm WR} = 1.5$. They found that for He star models with $f_{\rm WR} = 1.0$, those with final pre-SN masses between approximately 10 and $12 \,M_{\odot}$ exhibit relatively low compactness values. Similarly low compactness is also exhibited in our models for final pre-SN masses within this range (see Fig.~\ref{fig:compactness}).
Furthermore, \cite{ertl2020explosion} suggested that for He stars with final pre-SN masses between 6.5 and 12 $M_{\odot}$, the evolutionary outcome is a mixture of BHs and NSs, consistent with our results (see Fig.~\ref{fig:compactness}).
In this work, we did not consider BHs formed by fallback, resulting in no formation of low-mass BHs (see Fig.~\ref{fig:Gravitational mass}).

Previous studies demonstrated that the explodability of He stars exhibits non-monotonic behavior with initial mass \cite[e.g.,][]{ertl2020explosion,aguilera2023stripped}. However, these studies primarily explored He star explodability criteria and the impact of mass loss and metallicity on pre-SN core structure, which determine SN explosion properties such as remnant masses, nucleosynthesis (e.g., $^{56}$Ni yields), and light curves. In this work, we focused on the key physical mechanisms underlying the non-monotonic behavior. We found that the final core structure and explodability are mainly determined by $M_{\rm CO}$ and $X_{\rm C}$. These quantities served as the initial conditions for the subsequent burning stages, influencing the properties of central C/Ne burning and the locations of convective shells. In particular, the location of the last convective C-burning shell set the mass of the C-free core, which limits the iron core mass and the final compactness.

Notably, \cite{patton2020towards} inferred that models with $M_{\mathrm{CO}} > 10\, M_\odot$ will be very difficult to explode, based on the simulations of \cite{sukhbold2016core} and \cite{ertl2020explosion}. However, our models show that successful explosions can still occur at slightly higher $M_{\rm CO}$ values (see the black dashed vertical line in Fig.~\ref{fig:compactness}a). Specifically, models with $M_{\mathrm{He}}^{\mathrm{i}} = 22$ and $23\, M_\odot$ ($M_{\mathrm{CO}} = 10.30$ and $10.74\, M_\odot$, respectively) both yield successful explosions. This result is consistent with the recent study by \cite{Maltsev2025}, who also predicted a window of explodability for binary-stripped stars at solar metallicity, extending up to $M_{\rm CO}=13.2-15.4 \,M_{\odot}$. Significantly, the criterion of \cite{Maltsev2025} predicts a higher probability of explosion compared to that of the \cite{patton2020towards} and \cite{ertl2020explosion}.
This discrepancy likely arises from the treatments of input physics, including convective overshooting, the ${ }^{12} \mathrm{C}(\alpha, \gamma)^{16} \mathrm{O}$ reaction rate, mass-loss rate and metallicity \cite[e.g.,][]{sukhbold2014compactness,sukhbold2018high,woosley2019evolution,chieffi2020presupernova,aguilera2023stripped,temaj2024convective,Maltsev2025,Long2025RAA}.

We also found that these quantities (i.e., final compactness, iron core mass, central entropy, C-free core mass, and binding energy) all exhibit a similar non-monotonic trend with initial mass (see Fig. \ref{fig:compactness}), which
is consistent with previous studies on ZAMS stars  \cite[e.g.,][]{schneider2021pre,schneider2023bimodal,schneider2024pre,temaj2024convective,laplace2025s}. Note that the first compactness peak found in this work, at $M_{\rm CO}\approx 7.51\, M_{\odot}$, is higher than the corresponding peak in non-stripped stars \cite[e.g.,][]{schneider2023bimodal}.
This systematic change of the compactness peak toward higher CO core masses, also observed in stripped-envelope stars \cite[e.g., see Fig.~2 in][]{schneider2023bimodal} and other He star models \cite[e.g.,][]{woosley2019evolution,aguilera2023stripped,ertl2020explosion}.
This is because these models have a higher central C mass fraction at the end of core He burning than non-stripped stars with the same CO core mass.

\subsection{The impact of metallicity on pre-SN core structure} \label{sec:metallicity}

Stellar wind is an important process in massive star evolution, significantly affecting stellar structure and the final fate of a star \citep{langer2012presupernova,meynet2015impact,renzo2017systematic}. The mass loss rates of WR stars are strongly metallicity dependent \citep{crowther2007physical,hainich2014wolf,tramper2016new,sander2020nature}. Higher metallicity enhances radiation-driven winds by increasing the opacity of the envelope, resulting in stronger winds and elevated mass-loss rates \citep{maeder2000evolution,heger2003massive,langer2012presupernova,smith2014mass,groh2019grids}. 
To explore the impact of metallicity on core structure, we present the evolution of a $15\,M_{\odot}$ He star with different metallicities using Kippenhahn diagrams (see Fig.~\ref{fig:Z_kippen}). In these models,
during core He burning, those with higher metallicity exhibit core characteristics similar to those of lower-mass stars, namely, higher central density and lower central temperature. Additionally, as metallicity increases, enhanced stellar wind mass loss significantly reduces the mass of the convective He-burning core, thereby decreasing the mass of the CO core.
According to Eq.~\ref{eq:carbon_mass_fraction}, these changes result in a slight increase in $X_{\rm C}$ at the end of core He burning (see Table~\ref{physical parameters}), altering the nature of core C burning and triggering multiple convective C-burning shell episodes.

Specifically, in the $Z=0.0088$ model, core C burning begins radiatively. After core C depletion, a convective C-burning shell forms. The C-burning front stabilizes at $1.92\,M_{\odot}$, which is lower than the mass coordinate used for compactness evaluation, remaining at this position until core collapse. This results in a smaller C-free core and an expanded envelope, which limits the growth of the iron core and reduces its compactness (see Table \ref{physical parameters}). In the $Z = 0.02$ model (see the top-right panel of Fig.~\ref{fig:Z_kippen}), core C burning still beginning radiative, and two convective C-burning shell episodes are subsequently produced. The location of the last convective C-burning shell exceeds the mass coordinate used to evaluate compactness. As discussed in detail in Section~\ref{sec:structure}, the final compactness and iron core mass are relatively high, corresponding to a peak in compactness. In the $ Z = 0.03 $ mdoel (see the lower-left panel of Fig.~\ref{fig:Z_kippen}), central C-burning proceeds radiatively and is followed by two successive convective burning shell episodes. In contrast, the $Z = 0.04 $ model exhibits convective C-burning core (the lower-right panel of Fig.~\ref{fig:Z_kippen}), followed by three convective C-burning shell. The radiative core C burning in the $Z = 0.03$ model drives stronger core contraction, prompting earlier Ne ignition in the core and shifting the C-burning front inward to a lower mass coordinate ($\sim 2.10 \,M_{\odot} $) compared to the $Z = 0.04 $ model ($\sim 2.20 \,M_{\odot} $). In both the $Z = 0.03$ and $Z = 0.04$ models, the location of the last convective C-burning shell is lower than the mass coordinate used to evaluate compactness. This leads to a smaller iron core mass and reduced compactness (see Table \ref{physical parameters}). These results demonstrate that, although $M_{\rm CO}$ and $X_{\rm C}$ show distinct sensitivities to metallicity, their variations shape the pre-SN core structure and explodability. Similarly, \citet{Maltsev2025} also found that the islands of explodability exhibit a dependence on metallicity, shifting systematically toward higher $M_{\rm CO}$ as metallicity increases.

\subsection{The impact of convective overshooting on pre-SN core structure} \label{sec:overshooting}

\begin{figure}
	\includegraphics[trim=0 0 0 0, width=1\columnwidth]{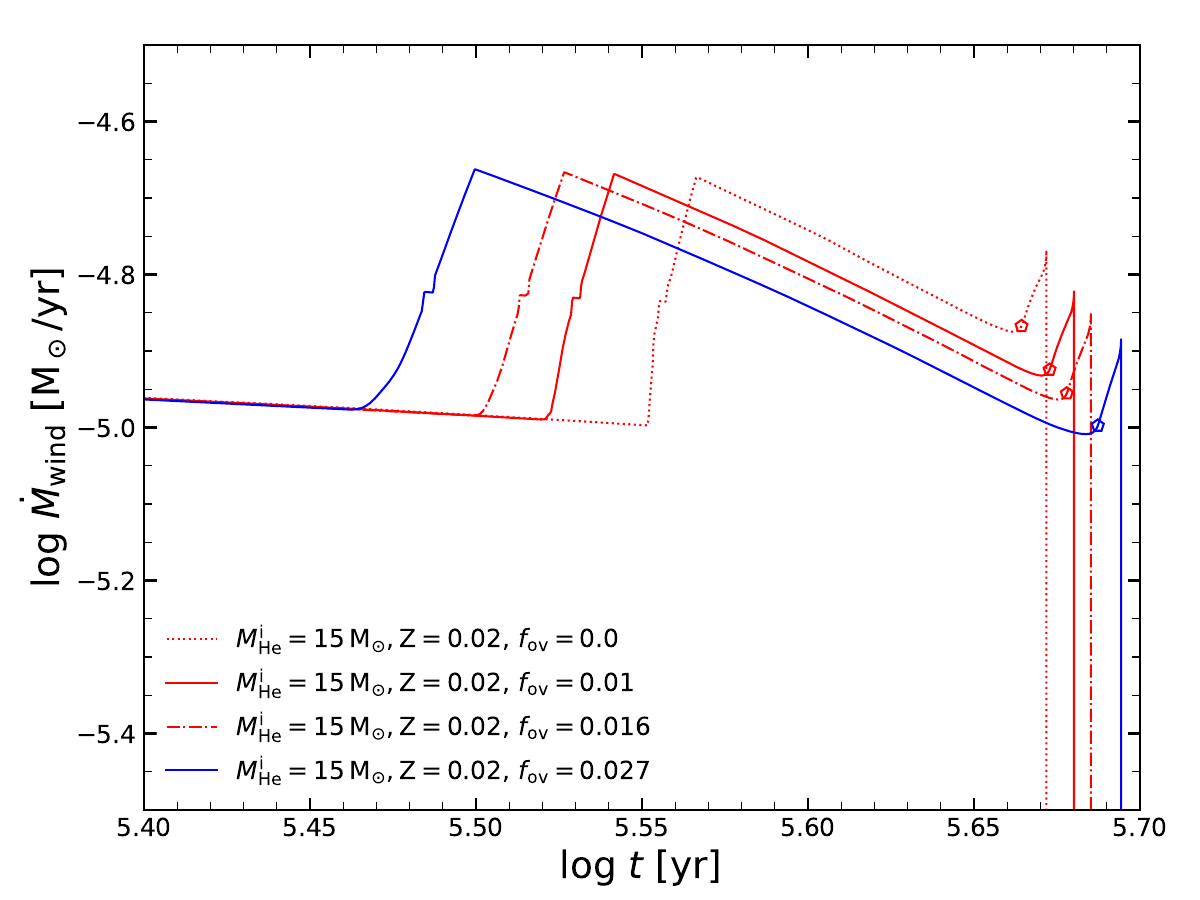}
	\caption{Rate of stellar wind mass loss as a function of evolutionary age for $M_{\mathrm{He}}^{\mathrm{i}} = 15 \,M_{\odot}$ models at $Z = 0.02$ with varying overshooting parameters. The red solid line is for $f_{\rm ov} = 0.01$, the blue solid line is for $f_{\rm ov} = 0.027$, the dash-dotted line is for $f_{\rm ov} = 0.02$, and the dotted line is for $f_{\rm ov} = 0.0$ (no overshooting). The pentagram indicates central He depletion.
		\label{fig:Mwind}}
\end{figure}
Convective overshooting transports fresh fuel to the burning regions while also moving nuclear reaction products outward. 
This process not only prolongs the lifetime of each burning phase but also alters the CO core mass and the mass fraction of ${}^{12}\rm C$ at the end of He burning, ultimately influencing the final pre-SN core structure \citep[e.g.,][]{langer2012presupernova,sukhbold2014compactness,kaiser2020relative, temaj2024convective}. However, significant uncertainties still exist regarding convection overshooting in He stars. For example, \cite{chanlaridis2022thermonuclear} and \cite{schneider2023bimodal} emphasized that variations in overshooting significantly influence the final core structure.

To explore the impact of convective overshooting on the pre-SN core structure, we considered four values of $f_{\rm ov} = 0.0, 0.01, 0.016, \text{and } 0.027$ \citep{herwig2000, kaiser2020relative, renzo2020predictions, higgins2021evolution, li2023grids}. Fig.~\ref{fig:overshooting} shows Kippenhahn diagrams for a $M_{\mathrm{He}}^{\mathrm{i}} =15\,M_{\odot}$ model with these convective overshooting values. In addition, Fig.~\ref{fig:Mwind} shows the stellar wind mass loss rate as a function of evolutionary age is displayed for the same models with varying overshooting values.
We found that He stars with larger overshooting paradoxically form smaller $M_{\rm CO}$ at central He exhaustion (see Table \ref{physical parameters}). This phenomenon arises from two competing effects: (1) overshooting mixes fresh He from the envelope into the core, prolonging the lifetime of core He burning. As shown in Fig.~\ref{fig:Mwind}, models with larger overshooting, such as $f_{\rm ov}=0.027$, have significantly longer central He-burning lifetimes compared to those with lower or no overshooting; and (2) overshooting also enhances the transport of core He burning products to the surface. Consequently, in models with larger overshooting, the surface helium abundance $Y_{\rm S}$ drops more rapidly below the WC/WO transition threshold ($Y_{\rm S} < 0.9$; see Section \ref{sec:mass-loss} ). The earlier transition from WNE to WC/WO winds (see Fig.~\ref{fig:Mwind}, where the blue solid line corresponds to the $f_{\rm ov}=0.027$ model that undergoes the earliest transition) triggers a sharp increase in the mass-loss rate ($\dot{M}_{\mathrm{WC}} \propto {Y_{\rm S}}^{0.44}$), which suppresses the growth of the CO core. Thus, enhanced overshooting reduces the CO core mass and increases the mass fraction of ${}^{12}\rm C$ in He stars. As shown in Table \ref{physical parameters}, when the overshooting parameter is increased from $f_{\rm ov}=0.0$ to 0.027, the CO core mass decreases progressively from 7.70 to 7.34 $M_{\odot}$, while $X_{\rm C}$ increases from 0.253 to 0.256. 
These small changes lead to a slight decrease in the location of the first convective C-burning shell (see Fig.~\ref{fig:overshooting}), thereby causing a reduction in both iron core mass and compactness, although the changes are not significant (see Table~\ref{physical parameters}).

\section{Conclusions} \label{sec:conclusion}
In this work, we investigated the evolution of He stars with initial masses from 5 to 65 $M_{\odot}$ at solar metallicity, focusing on their pre-SN core structures to predict their final fates. We also explored the effects of metallicity and convective boundary mixing on the pre-SN core structure. Our main conclusions are as follows:

(1) Several quantities exhibit similar non-monotonic behaviors with initial mass, such as the final compactness, iron core mass, central entropy, C-free core mass, and binding energy, reflecting their strong correlations. Based on the neutrino-driven SN model of \cite{muller2016simple} and the explodability criterion of \cite{ertl2016two}, we found that He stars with higher final compactness are more likely to form BHs. The final compactness peaks at initial masses of 15 $M_{\odot}$ and 30 $M_{\odot}$, corresponding to CO core masses of 7.51 $M_{\odot}$ and 13.98 $M_{\odot}$, respectively.

(2) By following the evolution of compactness with stellar age, we found that this parameter is governed not only by the nature of core C burning (i.e., convective or radiative) but also by the location and strength of the subsequent burning shells. Specifically, when the convective shell is located within the mass coordinate at which we evaluate the compactness, it significantly reduces the compactness. Notably, the location of the last convective C-burning shell is the dominant factor in determining the final compactness.

(3) We found that the final compactness increases with the transition of core C burning from the convective to the radiative regime ($M_{\mathrm{He}}^{\mathrm{i}}$ = 12--15$\,M_{\odot}$ in Fig.~\ref{fig:compactness}). This transition indicates that neutrino energy loss exceeds the energy released by core C burning, leading to stronger core contraction and further outward progression of the C-burning front, which yields a larger iron core mass and a higher final compactness. When core Ne burning becomes neutrino-dominated, the same mechanism causes the final compactness to gradually increase again ($M_{\mathrm{He}}^{\mathrm{i}}$ = 23--30$\,M_{\odot}$ in Fig.~\ref{fig:compactness}).

(4) After the compactness peaks, the final compactness decreases with increasing initial mass ($M_{\mathrm{He}}^{\mathrm{i}}$ = 15--18$\,M_{\odot}$ and 30--40$\,M_{\odot}$ in Fig.~\ref{fig:compactness}). We found that earlier core Ne/O ignition and shell mergers are key mechanisms driving this decrease. Specifically, when the location of the C-burning front (during the first convective C-burning shell) exceeds the effective Chandrasekhar mass, it triggers the earlier ignition of Ne and O, which slows core contraction. Moreover, when the entropy of the Ne/O-burning shells exceeds that of the layers above, a shell merger occurs, expanding the overlying layers and shifting the burning front to lower mass coordinates. These processes ultimately result in a smaller iron core and a lower final compactness.

(5) We found that metallicity and convective overshooting affect the central $^{12}\mathrm{C}$ mass fraction and CO core mass to different extents, leading to differences in the subsequent core structure. Specifically,
higher metallicity significantly decreases the CO core mass and slightly increases the central ${}^{12}\rm C$ mass fraction.
However, the final compactness does not follow a monotonic trend with metallicity, while the explodability is highly sensitive to metallicity variations. On the other hand, a larger convective overshooting also slightly increases the central ${}^{12}\rm C$ mass fraction but only slightly decreases the CO core mass. This leads to a monotonic decline in final compactness with increasing overshooting, but the explodability is not sensitive to changes in overshooting.
The influence of these uncertainties on the explodability of progenitors varies significantly.

\begin{acknowledgements}
This study is supported by the National Natural Science Foundation of China (Nos 12225304, 12288102, 12090040, 12090043, 12403035, 12173010, 12573034, 12273105), the National Key R\&D Program of China (No. 2021YFA1600404), the Yunnan Revitalization Talent Support Program (Yunling Scholar Project and Innovation Team Project), the Youth Innovation Promotion Association CAS (No. 2021058), the Yunnan Fundamental Research Project (Nos. 202501AS070005, 202401AV070006, 202501AS070078),  ,the International Centre of Supernovae (ICESUN), Yunnan Key Laboratory (No. 202505AV340004), the Postdoctoral Fellowship Program of CPSF (No. GZB20240307), the China Postdoctoral Science Foundation (Nos. 2024M751375, 2024T170393), and the Jiangsu Funding Programme for Excellent Postdoctoral Talent (No. 2024ZB705).
\end{acknowledgements}

\bibliographystyle{aa} % style aa.bst
\bibliography{paper} % your references Yourfile.bib

\onecolumn
\begin{appendix}
	
\section{Detailed model properties}
	
\begin{table}[h!]
	\caption{Initial configurations and pre-SN properties of our default models.}\label{parameter}
	\begin{center}
		\begin{tabular}{
				l@{\hspace{4mm}}c@{\hspace{4mm}}c@{\hspace{4mm}}
				c@{\hspace{4mm}}c@{\hspace{4mm}}c@{\hspace{4mm}}
				c@{\hspace{4mm}}c@{\hspace{4mm}}c@{\hspace{4mm}}
				c@{\hspace{4mm}}c@{\hspace{4mm}}c@{\hspace{4mm}}
				c@{\hspace{4mm}}c@{\hspace{4mm}}c@{\hspace{4mm}}
				c@{\hspace{4mm}}c@{\hspace{4mm}}c@{\hspace{4mm}}
			}
			\hline\hline
			
			$M_{\text{He}}^{\text{i}}$ & $M_{\rm CO}$ & $X_{\text{C}}$ & $M_{\text{C-free}}$ & $\xi_{2.5}$ & $\mu_{4}$ & $M_{4}$ & $s_{\text{c}}$ & $M_{\rm f}$ & $M_{\rm Fe}$ & log$(-E_{\text{bind}})$ & $E_{\text{exp}}$ & $\text{Fate}$ & $M_{\text{rm,grav}}$
			\\
			$\left[M_{\odot}\right]$ & $\left[M_{\odot}\right]$ &  &  $\left[M_{\odot}\right]$   &    &  &  &$\left[N_{\mathrm{A}} k_{\mathrm{B}}\right] $&$\left[M_{\odot}\right]$& $\left[M_{\odot}\right]$&$\left[\text{erg}\right] $ & $\left[10^{51}\text{erg}\right] $  & &$\left[M_{\odot}\right]$ \\
			\hline
			5.0 & 2.44 & 0.355 & 1.71 & 0.063 & 0.03 & 1.61 & 0.84 & 4.00 & 1.51 & 50.95 & 0.52 & NS & 1.41 \\
			6.0 & 3.01 & 0.338 & 1.87 & 0.131 & 0.06 & 1.70 & 0.88 & 4.66 & 1.52 & 51.14 & 0.75 & NS & 1.69 \\
			7.0 & 3.58 & 0.324 & 1.82 & 0.116 & 0.04 & 1.67 & 0.87 & 5.30 & 1.53 & 51.12 & 0.44 & NS & 1.46 \\
			8.0 & 4.14 & 0.311 & 1.56 & 0.177 & 0.08 & 1.58 & 0.85 & 5.90 & 1.49 & 51.28 & 0.82 & NS & 1.40 \\
			9.0 & 4.70 & 0.300 & 1.86 & 0.198 & 0.06 & 1.79 & 0.93 & 6.48 & 1.61 & 51.32 & 0.51 & NS & 1.55 \\
			10.0 & 5.25 & 0.290 & 2.54 & 0.330 & 0.10 & 2.00 & 1.02 & 7.05 & 1.65 & 51.49 & $\cdots$ & BH & 7.05 \\
			11.0 & 5.81 & 0.281 & 2.23 & 0.233 & 0.08 & 1.80 & 0.95 & 7.59 & 1.62 & 51.43 & 0.86 & NS & 1.57 \\
			12.0 & 6.35 & 0.273 & 1.52 & 0.117 & 0.07 & 1.52 & 0.82 & 7.84 & 1.48 & 51.29 & 0.60 & NS & 1.39 \\
			13.0 & 6.88 & 0.266 & 2.20 & 0.244 & 0.09 & 1.92 & 0.99 & 8.06 & 1.66 & 51.44 & 1.01 & NS & 1.67 \\
			14.0 & 7.27 & 0.259 & 2.66 & 0.352 & 0.11 & 2.01 & 1.03 & 8.41 & 1.73 & 51.54 & 1.80 & NS & 1.81 \\
			15.0 & 7.51 & 0.254 & 3.28 & 0.571 & 0.16 & 2.32 & 1.14 & 8.79 & 1.85 & 51.70 & $\cdots$ & BH & 8.79 \\
			16.0 & 7.82 & 0.249 & 2.84 & 0.541 & 0.16 & 2.30 & 1.09 & 9.20 & 1.77 & 51.69 & $\cdots$ & BH & 9.20 \\
			17.0 & 8.18 & 0.244 & 2.72 & 0.459 & 0.14 & 2.17 & 1.08 & 9.63 & 1.74 & 51.70 & $\cdots$ & BH & 9.63 \\
			18.0 & 8.57 & 0.240 & 1.83 & 0.198 & 0.06 & 1.79 & 0.93 & 10.05 & 1.60 & 51.55 & $\cdots$ & BH & 10.05 \\
			19.0 & 8.98 & 0.236 & 2.01 & 0.202 & 0.07 & 1.69 & 0.90 & 10.47 & 1.56 & 51.57 & 0.16 & NS & 1.47 \\
			20.0 & 9.41 & 0.231 & 1.98 & 0.167 & 0.06 & 1.72 & 0.90 & 10.90 & 1.57 & 51.52 & 0.35 & NS & 1.51 \\
			21.0 & 9.86 & 0.227 & 2.08 & 0.188 & 0.07 & 1.81 & 0.93 & 11.33 & 1.57 & 51.58 & 0.44 & NS & 1.57 \\
			22.0 & 10.30 & 0.223 & 2.19 & 0.235 & 0.09 & 1.92 & 0.97 & 11.76 & 1.64 & 51.62 & 0.61 & NS & 1.68 \\
			23.0 & 10.74 & 0.220 & 2.29 & 0.220 & 0.08 & 1.81 & 0.96 & 12.20 & 1.62 & 51.63 & 0.55 & NS & 1.58 \\
			24.0 & 11.19 & 0.216 & 2.31 & 0.354 & 0.12 & 2.08 & 1.03 & 12.65 & 1.70 & 51.72 & $\cdots$ & BH & 12.65 \\
			25.0 & 11.64 & 0.212 & 2.40 & 0.439 & 0.13 & 2.20 & 1.05 & 13.11 & 1.74 & 51.75 & $\cdots$ & BH & 13.11 \\
			26.0 & 12.10 & 0.209 & 2.57 & 0.477 & 0.13 & 2.28 & 1.06 & 13.56 & 1.77 & 51.78 & $\cdots$ & BH & 13.56 \\
			27.0 & 12.56 & 0.206 & 2.54 & 0.527 & 0.14 & 2.33 & 1.08 & 14.03 & 1.79 & 51.82 & $\cdots$ & BH & 14.03 \\
			28.0 & 13.02 & 0.202 & 2.61 & 0.548 & 0.15 & 2.35 & 1.10 & 14.49 & 1.77 & 51.85 & $\cdots$ & BH & 14.49 \\
			29.0 & 13.50 & 0.199 & 2.70 & 0.580 & 0.20 & 2.32 & 1.10 & 14.96 & 1.83 & 51.88 & $\cdots$ & BH & 14.96 \\
			30.0 & 13.98 & 0.196 & 2.79 & 0.609 & 0.22 & 2.31 & 1.10 & 15.44 & 1.86 & 51.90 & $\cdots$ & BH & 15.44 \\
			35.0 & 16.39 & 0.182 & 3.37 & 0.607 & 0.29 & 2.05 & 1.13 & 17.85 & 1.86 & 52.05 & $\cdots$ & BH & 17.85 \\
			40.0 & 18.88 & 0.170 & 3.87 & 0.365 & 0.12 & 1.82 & 0.99 & 20.33 & 1.69 & 52.06 & $\cdots$ & BH & 20.33 \\
			45.0 & 21.39 & 0.159 & 4.46 & 0.420 & 0.12 & 2.05 & 1.07 & 22.85 & 1.83 & 52.16 & $\cdots$ & BH & 22.85 \\
			50.0 & 23.96 & 0.150 & 5.01 & 0.469 & 0.13 & 2.12 & 1.12 & 25.43 & 1.87 & 52.26 & $\cdots$ & BH & 25.43 \\
			55.0 & 26.54 & 0.141 & 5.70 & 0.595 & 0.15 & 2.32 & 1.19 & 28.03 & 1.95 & 52.35 & $\cdots$ & BH & 28.03 \\
			60.0 & 29.14 & 0.134 & 6.17 & 0.777 & 0.19 & 2.49 & 1.29 & 30.65 & 2.07 & 52.45 & $\cdots$ & BH & 30.65 \\
			65.0 & 31.77 & 0.127 & 6.63 & 0.808 & 0.22 & 2.64 & 1.31 & 33.30 & 2.09 & 52.46 & $\cdots$ & BH & 33.30 \\
			\hline			    	    			    		    	    	    	    	    	    
			\hline
		\end{tabular}
	\end{center}
	{\bf Notes:} The table summarizes key pre-SN properties for each He star model with the same initial parameters, including MESA's default ${}^{12} \mathrm{C}(\alpha, \gamma)^{16} \mathrm{O}$ rate from the NACRE compilation \citep{angulo1999compilation}, an initial metallicity of $Z = 0.02$, and a convective overshooting parameter of $f_{\rm ov} = 0.01$.
	Column (1): Initial He star mass, $M_{\text{He}}^{\text{i}}$, (the $65\,M_{\odot}$ model experiences a weak PPISN);
	Column (2): CO core mass, $M_{\rm CO}$, defined as the maximum mass coordinate where the central ${ }^4 \mathrm{He}$ mass fraction, $X_{\rm He}$, drops below ${10}^{-1}$ at the end of core He-burning, following \cite{tauris2015ultra};
	Column (3): Central ${}^{12}\rm C$ mass fraction, $X_{\rm C}$, at the end of core He burning, defined as $X_{\rm He}<{10}^{-4}$;
	Column (4): Carbon-free core mass, $M_{\rm C-free}$, defined as the maximum mass coordinate where the central ${}^{12}\rm C$ mass fraction, $X_{\rm C}$, is less than ${10}^{-5}$;
	Column (5): Compactness, $\xi_{2.5}$, (defined by Eq.\,\ref{xi});
	Column (6): Radial mass derivative, $\mu_{4}$, (as defined in Eq.\,\ref{mu_4});
	Column (7): Mass coordinate, $M_{4}$, at a specific entropy of $s = 4$;
	Column (8): Central specific entropy, $s_{\rm c}$, at the onset of iron-core collapse;
	Column (9): Final pre-SN mass, $M_{\rm f}$;
	Column (10): Iron core mass, $M_{\rm Fe}$;
	Column (11): Gravitational binding energy, $-E_{\rm {bind}}$, of the material above the iron core \citep[see also][]{temaj2024convective};
	Column (12): Explosion energy $E_{\rm {exp}}$;
	Column (13): Predicted final fate of the star;
	Column (14): Gravitational compact remnant mass, $M_{\rm {rm,\,grav}}$. Both the predicted
	final fate, $E_{\rm {exp}}$, and $M_{\rm {rm,\,grav}}$ are derived from the semi-analytic model developed by \cite{muller2016simple}.
\end{table}

\clearpage
\section{The impact of metallicity on pre-SN core structure}

\begin{figure}[h!] 
	\centering
	\includegraphics[width=1\textwidth]{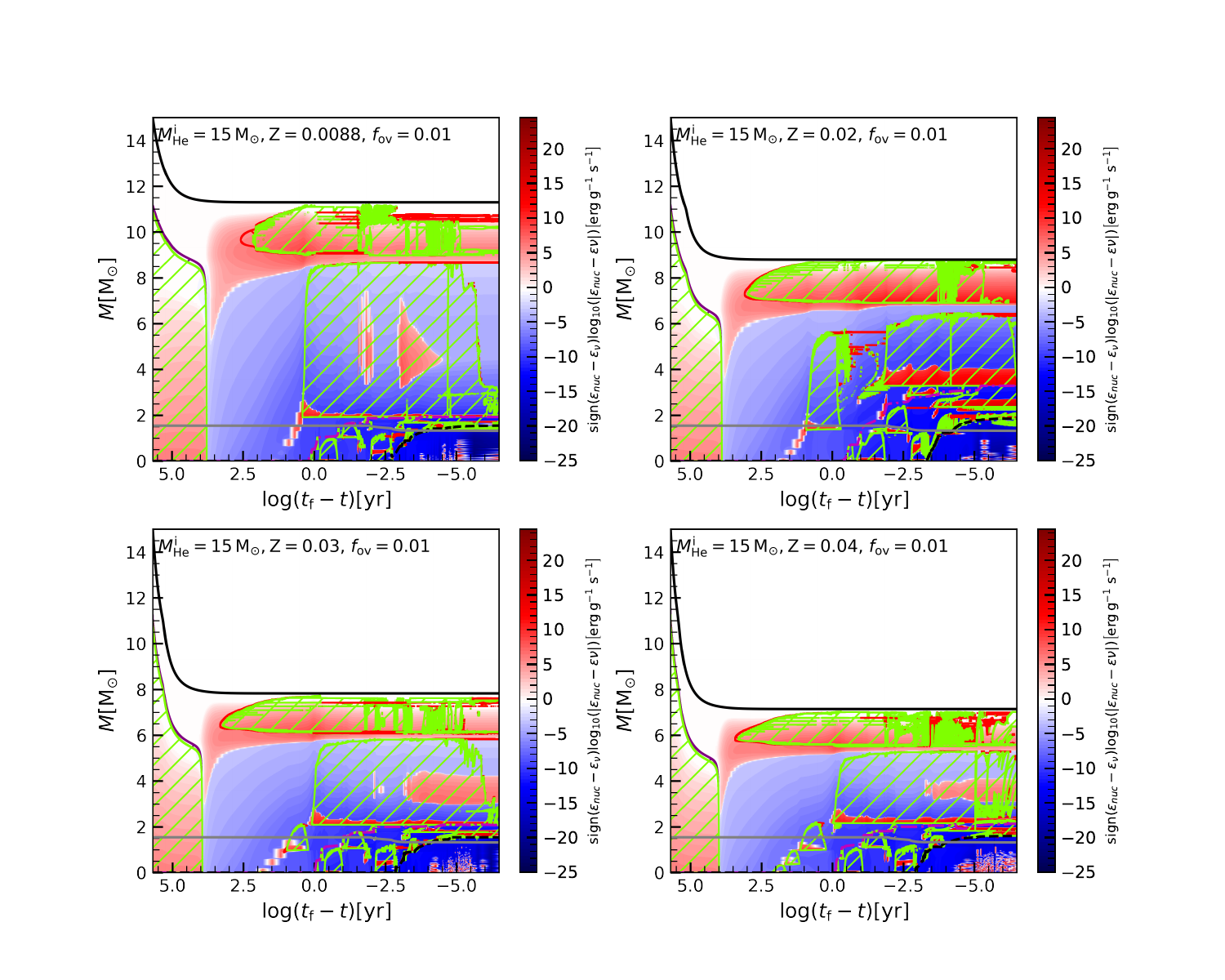}
	\caption{Same as Fig.~\ref{fig:xi}, but for a \(M_{\mathrm{He}}^{\mathrm{i}} = 15\, M_{\odot}\) model with varying metallicities: upper-left panel (\(Z = 0.0088\)), upper-right panel (\(Z = 0.02\)), lower-left panel (\(Z = 0.03\)), and lower-right panel (\(Z = 0.04\)).}
	\label{fig:Z_kippen}
\end{figure}

\clearpage
\section{The impact of convective overshooting on pre-SN core structure}

\begin{figure}[h!]
	\centering
	\includegraphics[width=1\textwidth]{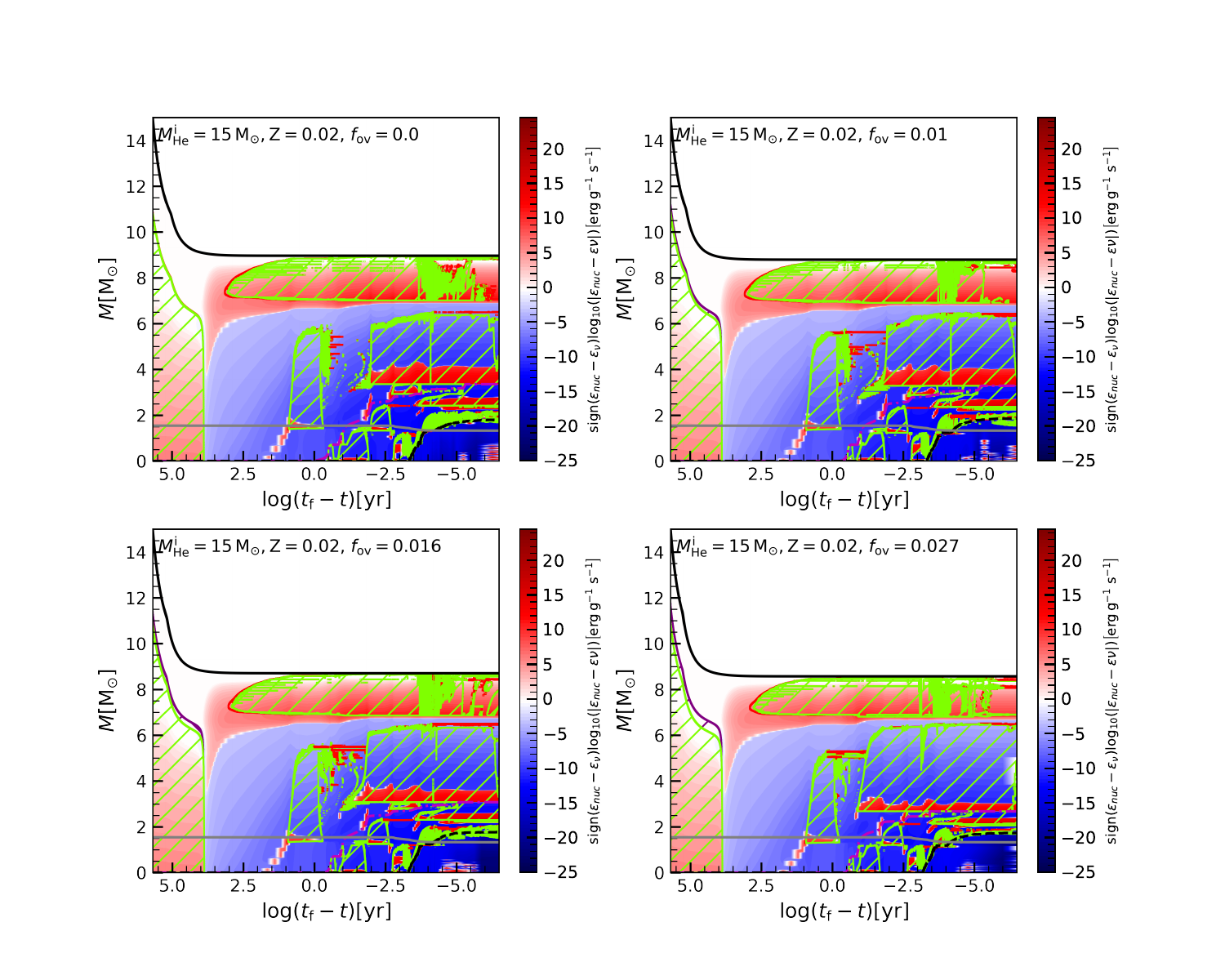}
	\caption{Same as Fig.~\ref{fig:xi}, but for a \(M_{\mathrm{He}}^{\mathrm{i}} = 15\,M_{\odot}\) models with varying overshooting: upper-left panel ($f_{\rm ov} = 0.0$), upper-right panel ($f_{\rm ov} = 0.01$), lower-left panel ($f_{\rm ov} = 0.016$) and lower-right panel ($f_{\rm ov} = 0.027$).}
	\label{fig:overshooting}
\end{figure}

\clearpage
\section{Electron degeneracy parameter} \label{sec:Electron degeneracy}
	
\begin{figure}[h!]  
    \centering
    \scalebox{1}{
      \begin{tabular}{ccc}
        \includegraphics[trim = 0 0 0 0 ,width=1\textwidth]{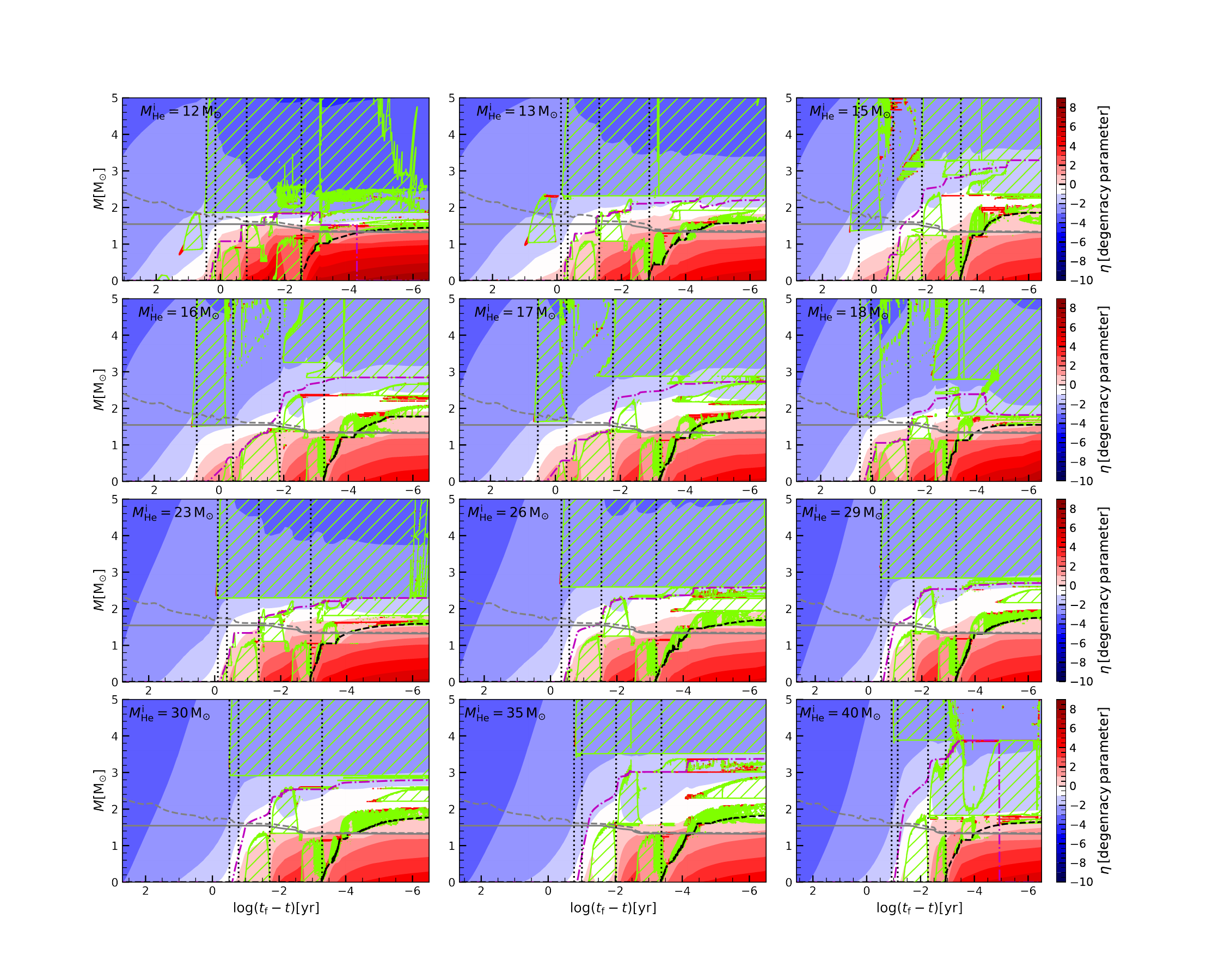}
      \end{tabular}%
    }
    \caption{Same as Fig. \ref{fig:11-18kippen}, but the color bar represents the dimensionless electron degeneracy parameter, as described in panel (d) of Fig. \ref{fig:xi}.}
	\label{fig:He_12-18_eta_kippen}
  \end{figure}%

\end{appendix}

\end{document}